\newif\ifShowKeys
\ShowKeystrue

\documentclass[11pt,a4paper]{article} 
\pdfoutput=1
\usepackage[no-natbib-sort]{my-jheppub}
\usepackage{amsmath, amssymb}

\usepackage{mathpazo}
\usepackage{bm}
\usepackage{environ}
\usepackage{mathrsfs}
\usepackage{array,arydshln}

\usepackage{graphicx,epsfig}
\usepackage{epic}
\usepackage{youngtab}
\usepackage{float}
\usepackage{color}
\definecolor{maroon}{rgb}{0.8,0.3,0.}

\usepackage{slashed}
\usepackage[nodayofweek]{date time}

\usepackage{hyperref}

\usepackage{aurical}
\usepackage[T1]{fontenc}

\newcommand \foot [1] {\footnote{#1\vspace{2pt}}}
\newcommand \rf [1] {(\ref{#1})}

\usepackage{framed}
\definecolor{shadecolor}{RGB}{255, 230, 204}

\allowdisplaybreaks

\newmuskip\pFqmuskip

\newcommand*\pFq[6][8]{%
  \begingroup % only local assignments
  \pFqmuskip=#1mu\relax
  \mathcode`\,=\string"8000
  \begingroup\lccode`\~=`\,
  \lowercase{\endgroup\let~}\pFqcomma
  {}_{#2}F_{#3}{\big[\genfrac..{0pt}{}{#4}{#5};#6\big]}%
  \endgroup
}

\newcommand*\pFtildeq[6][8]{%
  \begingroup % only local assignments
  \pFqmuskip=#1mu\relax
  \mathcode`\,=\string"8000
  \begingroup\lccode`\~=`\,
  \lowercase{\endgroup\let~}\pFqcomma
  {}_{#2}\widetilde{F}_{#3}{\big[\genfrac..{0pt}{}{#4}{#5};#6\big]}%
  \endgroup
}

\newcommand{\pFqcomma}{\mskip\pFqmuskip}

\newcommand{\be}{\begin{equation}}
\newcommand{\ee}{\end{equation}}

\newcommand{\mc}{\mathcal }

\newcommand{\bv}{\begin{verbatim}}
\newcommand{\ev}{\end{verbatim}}

\newcommand{\la}{\label}

\newcommand{\eps}{\epsilon}

\newcommand{\rs}{\textsf{\textbf s}}
\newcommand{\rt}{\textsf{\textbf t}}
\newcommand{\ru}{\textsf{\textbf u}}

\title{On triviality of S-matrix\\ in  conformal higher spin theory
}
\author[a]{Matteo Beccaria,\footnote{matteo.beccaria@le.infn.it}} 
\author[b]{Simon Nakach\footnote{simon.nakach09@imperial.ac.uk}} 
\author[b]{ and \ Arkady A. Tseytlin\footnote{Also at Lebedev Institute, Moscow. \ \ tseytlin@imperial.ac.uk}} 

\abstract{
We consider the conformal higher spin (CHS) theory in $d=4$ that  contains  the $s=1$ Maxwell  vector, 
$s=2$  Weyl graviton and their higher spin 
$s=3,4,...$ counterparts  with higher-derivative $\Box^s$   kinetic terms. The  interacting action  for such theory 
can be  found as the coefficient of the logarithmically  divergent  part in the induced action for sources coupled to 
higher spin currents in   a free complex scalar field model. 
We explicitly determine   some cubic and quartic interaction vertices 
in the CHS action from scalar loop  integrals. We then compute the simplest  tree-level 4-particle  scattering amplitudes 
11$\to$11, 22$\to$22 and  11$\to$22  and find that after  summing up  all the  intermediate CHS  exchanges they vanish. 
This generalises %the vanishing  of the  graviton amplitude in Weyl theory and  also 
 the vanishing of the   scattering  amplitude for external conformal scalars  interacting via the 
 exchange of  all  CHS fields    found  earlier in   arXiv:1512.08896. 
 This   vanishing  should generalise  to all scattering amplitudes in the CHS theory   and 
 as in the conformal scalar  scattering case 
 should be a consequence of the underlying infinite dimensional  higher spin  symmetry
 that extends the standard conformal symmetry. 
}

\affiliation[a]{Dipartimento di Matematica e Fisica Ennio De Giorgi,\\
Universit\`a del Salento \& INFN, Via Arnesano, 73100 Lecce, 
Italy} 

\affiliation[b]{The Blackett Laboratory, Imperial College, London SW7 2AZ, U.K.}

\allowdisplaybreaks

\begin{document}

\date{\currenttime}

%   \begin{flushleft}\boxed{\small{\tt \today \ \ - \ \  \currenttime }}\end{flushleft}

 \begin{flushright}\small{Imperial-TP-AT-2016-{02}}\end{flushright}				% report number

\maketitle
\flushbottom

\def \ci {\cite} \def \ed {\end{document}}\def \foot {\footnote}\def \del {\partial}\def \vp {\varphi}

\def \la {\label}
\def \fo {{\textstyle { 1 \ov 4}}}
\def \la {\label}
\def \str {{\rm str}}
\def \Tr {{\rm Tr}}
\def \ha {{ \textstyle{1 \over 2}}}
\def \td {\tilde}
\def \ci{\cite}
\def \del{ \partial}
\def\ov{\over}
\def\no{\nonumber} 
\def \aa {{\rm a}}
\def \p {\phi}
\def \m {\mu}\def \n {\nu} 
\def \te {\textstyle} 
\def \ed  {\end{document} }
\def \fo {{ \textstyle{1 \over 4}}}
\def \be {\begin{equation}}
\def \ee {\end{equation}}
\def \ed {\end{document}}
\def \ba {\begin{align}}
\def \era {\end{align}}
\def \a {\alpha}
\def \b {\beta}
\def \g {\gamma} 
\def \l {\lambda}
\def\e{\epsilon}

\def \bi{\bibitem}
\def \la {\label}

\def \l {\lambda}
\def\foot{\footnote}
\def \tl  {{\tilde \l}}
\def \sql {{\sqrt \l}}
\def \adss {$AdS_5 \times S^5$\ }

\def \ov {\over}
\def \rc {{\rm k}}
\def \n {\nu} \def \m {\mu}\def \G  {\Gamma} \def \J  {{\cal J}}

\def \kg {{g}}
\def \l {\lambda} \def \r {\rho}
\def \V  {{\rm V}}
 \def \sqt {{\te {1\ov \sqrt 2}}}
 \def \VV {{\rm V}}
 \def \P {\Pi}
\def \sym {{\rm sym}}
\def \SS {S}
\def \F  {{\rm F}}
\def \s  {\sigma}
%%%%%%%%%%%%%%%%%%%%%%%%%%
\def \iffa  {\iffalse} 
\def \j  {{\rm j}}
\def \V {{\rm V}} 
\def \ie  {{\em i.e.\ }}
\def \h {\eta} 
\def \aa {{a}} 

%%%%%%%%%%%%%%%%%%%%%%%%%%%%%%%
\section{Introduction}
%%%%%%%%%%%%%%%%%%%%%%%%%%%%%%%%%
Conformal higher spin (CHS)   theory   \ci{Fradkin:1985am,Fradkin:1989md,Tseytlin:2002gz,Segal:2002gd,Bekaert:2010ky,Beccaria:2014jxa,Haehnel:2016mlb} 
is a remarkable example  of a  formally  consistent (gauge-invariant, though higher-derivative and thus non-unitary) 
 higher spin  model   that has a local action  with a flat-space vacuum and one  dimensionless coupling constant. 
 
 It is naturally associated with another consistent higher spin theory, {\em i.e.} 2-derivative massless   higher spin theory 
 in AdS space of one dimension higher. Starting with a free complex scalar equation in 4  dimensions\foot{In this paper we shall concentrate on the $d=4$ case  but most of our  discussion may  be generalised to even $d>4$.}
 $\del^2 \vp=0$  one  gets an infinite tower of conserved  traceless  totally symmetric higher spin currents
  $J_{s}=  \vp^* \J_s \vp \ , \ \ \J_s \sim   \del^s  + ...$, \ $s=1,2,...,\infty$ that generalise the spin 0 
   primary operator $J_0 = \vp^*\vp$.  
 % \comment\red{the case $s=0$ is not a conserved current, agree ? }
    %They are dual to massless higher  spins in AdS  while 
  The conserved  charges  generate an  infinite dimensional symmetry algebra  of the free scalar equation \ci{Eastwood:2002su,Vasiliev:2003ev} 
  that is associated to a collection of   conformal Killing tensors.  The CHS theory may be viewed as a gauge theory of this  higher spin global symmetry. 
  A closely related approach is based on  interpreting 
     CHS fields   as "sources"   for the currents  $J_{s}$ that then inherit  the  linearised gauge invariances 
  $\delta  h_{\m_1...\m_s} =\del_{(\m_1 } \epsilon_{\m_2 ... \m_s)} + \eta_{(\m_1\m_2} \a_{\m_3...\m_s)}$ 
which   generalise the  usual reparametrisations and Weyl symmetry of  conformal gravity. 
  Starting with  free  $U(N)$  complex  scalar   CFT,  adding source terms $h_s J_s$ to the  free action $\del \vp^* \del \vp$ 
  and integrating over  $\vp$  one finds for  the generating  functional 
  of correlators of the currents $J_s$ 
   \be \la{1}
	\G[h]= N \,\log \det \big( \del^2   + \sum_s h_s\,\J_s\big)\ ,\qquad \ \ \ \ \quad \J_s \sim \del^s\ . 
\ee
From the vectorial AdS/CFT point of view  \ci{Klebanov:2002ja}    the 4d currents $J_s$ are dual to  massless   higher spins in AdS$_5$ 
and the generating functional $\G[h]$    should  then be  equal to the on-shell value of the AdS  action with $h_s$ 
being the boundary values  of the higher spin fields in AdS. 

One  can   then obtain  a local gauge-invariant action for the CHS fields $h_s$   by  identifying  it  with the 
 logarithmically UV divergent  part  of the "induced" action \rf{1} \ci{Tseytlin:2002gz,Segal:2002gd,Bekaert:2010ky}
 \ba
& S[h] \sim  \,\log \det \big( \del^2   + \sum_s h_s\,\J_s\big)\Big|_{\log \Lambda} \no\\
&\quad  \sim 
 { 1 \ov \kg^2}   \sum_s
  \int  d^4 x \Big(  h_s  \del^{2s } h_s    +     \del^{s_1+s_2+s_3 - 2    }  h_{s_1}  h_{s_2}  h_{s_3}     
+   \del^{ s_1+s_2+s_3 + s_4 - 4    }   h_{s_1}  h_{s_2}   h_{s_3} h_{s_4} + ...   \Big)  \ . 
\la{2}
 \end{align}
Here we introduced an arbitrary dimensionless coupling constant $\kg$  and  
indicated  symbolically   the  overall powers of derivatives 
in the  kinetic and interaction terms  that  follow from dimensional analysis. 
Indeed, as the  4d scalar $\vp$  has mass dimension 1, the   current $J_s$   has dimension $2+s$ 
and thus  the corresponding source field $h_s$  must have the "shadow"-field dimension $\Delta_s= 2-s $
(i.e.   1  for vector  field, 0 for conformal  graviton, etc.). This then determines the derivative structure of  \rf{2}.\foot{
The   
 fact  that the powers of derivatives  are directly correlated with  the values of the  spins in the vertex 
 (which is a  consequence of the underlying  conformal invariance) is an important simplifying feature  of this theory 
 compared to the AdS  higher spin theory and a  hypothetic   2-derivative massless  higher spin theory  in flat space 
 that both contain a dimensional parameter.}
 In particular, the presence of $2s$ derivatives in the kinetic term in \rf{2} is  consistent  
 with both  the  above  linearised gauge invariance $\delta h_s = \del \e_{s-1}   + \eta_2 \a_{s-2}$   and  the locality 
 of the action.\foot{The kinetic term   should   contain the transverse traceless spin $s$  projector 
 $\Pi_s$   that  is given by  products of $s$   factors of $\Pi^\m_\n = \delta^\m_\n - {\del^\m \del_\n\ov \del^2}$ 
 and thus $\Pi_s \del^{2s}$   is local.}

In addition  to the  linearised   gauge symmetry and the standard conformal symmetry  
the CHS action should  be invariant   under   the full infinite dimensional CHS  symmetry 
\ci{Segal:2002gd}    whose   global part  is the symmetry of  free scalar 4d Laplace  equation. 
This  large    symmetry should  %  lead to important simplification
provide   strong constraints  on the corresponding classical and quantum theory. 
 For example,  for fixed  spectrum of the  CHS  fields 
the action \rf{2}  should be  essentially   unique (modulo field redefinitions)\foot{Starting  instead 
from  a free scalar 
CFT  with a free   spinor  or free Maxwell  vector  CFT  one gets  a different  spectrum  of  conserved currents 
and thus a different  "induced"  CHS theory.} 
  and thus   renormalisable.
 In view of  the conformal symmetry being   gauged  here  it should  actually be 
  UV finite,  
provided  the theory is quantum-consistent, \ie  there are no  conformal and higher symmetry  anomalies.
An   indication  of  a  hidden simplicity of the  CHS  theory  is the  vanishing of the regularised 
total number of its degrees of freedom, or, equivalently, the triviality of the free (one-loop) partition function in flat space 
\ci{Beccaria:2015vaa}. This  partition  function   vanishes 
also on  4-sphere    implying the vanishing of the 
 Weyl anomaly a-coefficient \ci{Giombi:2013yva,Tseytlin:2013jya}  (which  is also in agreement with a  triviality of
 the  1-loop correction to the 
massless  HS partition function   as required by the  AdS/CFT \ci{Beccaria:2014xda}).\foot{The definition of the 
sum over spins requires 
a particular prescription that should be consistent with the underlying symmetries 
  \ci{Giombi:2014iua,Beccaria:2015vaa}.}
   Similar     vanishing was  also found (under some natural assumptions) 
         for the 1-loop   Weyl  anomaly 
  $c$-coefficient  \cite{Tseytlin:2013jya,Giombi:2014iua,Beccaria:2014xda,Beccaria:2015vaa}.
  As the Weyl   symmetry is one of the   CHS  gauge 
 symmetries,  the same anomaly  cancellation  may apply also to all  algebraic  CHS  gauge symmetries. 
 % under the assumption 
%  that contributions  to conformal anomaly from higher derivative   CHS  operators 
%  on Ricci flat background  factorise.

 The global part of the  CHS     symmetry  should  also   strongly  constrain other "observables",  {\em e.g.}, 
  the  analog of the   S-matrix   involving exchanges of the CHS fields.
  Indeed, it was found in   \ci{Joung:2015eny}  that  starting with  a free   external scalar field   coupled   (via 
  the above  current    $\int d^4 x\  h_s J_s  $ interaction) to  free  CHS fields with the action 
  $ \int  d^4 x \sum_s  h_s\,  \Pi_s  \,  \del^{2s } h_s $  and  
   computing the 4-scalar tree level scattering   due to the exchange of the tower of 
   CHS fields   one finds  that while  the individual   spin $s$ exchange   contributions are non-trivial, their sum over all $s=0,1,2,...$ 
  vanishes.
    This vanishing can be understood  \ci{Joung:2015eny} as  a consequence  of 
  the  CHS global symmetry  of the  coupled  theory (in particular,  the 
  "hypertranslations" $\delta \vp = \e^{\m_1...\m_s}  \del_{\m_1} ...\del_{\m_s}  \vp$ and scale invariance). 
  
  The  aim of the present paper is to  show that this triviality of the 4-particle scattering amplitude is found also % in the 
 % interacting CHS theory itself, {\em i.e.}  
   when  the external scalars are replaced by the CHS fields  themselves 
      with cubic and quartic interactions given by \rf{2}. We shall consider a few particular   examples  of  the CHS  4-particle scattering amplitudes (4-vector, 4-graviton, etc.) 
  and  find that after summation over all  exchanged   conformal higher spins 
  the total   amplitude vanishes. This   cancellation is  rather  non-trivial  and %should  %have a  reason behind it. 
   like in the external scalar  amplitude  case  \ci{Joung:2015eny}     should  again  be 
  a consequence of the underlying higher spin global symmetry of the theory  (and 
  should    thus  be a manifestation 
  of  a "generalised"  Coleman-Mandula theorem).
 %  \comment{should we mention here difficulties with pole cancellation with external higher derivative propagators ? }}
This   suggests    that the full  "S-matrix"  of the CHS theory should be trivial.\foot{As we are dealing with a  non-unitary higher derivative theory containing  
   an infinite number of  fields some  assumptions  of the  standard Coleman-Mandula theorem  may not directly  apply.
   %(in particular, the standard  definition of S-matrix  may not apply due to  non-standard
   In particular, the   definition of  the scattering matrix    for higher-derivative  fields  requires    clarification, see below.}

To be able to compute   scattering   amplitudes  of CHS states one needs   first to determine the precise 
structure of vertices in the  "induced" action \rf{2}. For that one   needs to find   the logarithmically divergent
 (or $1\ov \varepsilon$ pole   in dimensional regularisation) terms in the one-loop scalar  loop diagrams   with the two, three or four 
 current $J_s$  insertions. We shall   choose the external $h_s$ legs to be in the transverse  traceless   gauge.\foot{This 
   avoids, in particular,  the explicit discussion of  field redefinitions eliminating the traces.} 
   Having found the relevant terms in the action \rf{2}   we  will define the 4-particle scattering S-matrix 
   as the  amputated tree-level Green's function ({\em i.e.}  
    the sum of the exchange term and  contact vertex >\!----\!< + >\!<) 
    %\comment\red{is my version of exchange
   % plus contact picture ok ? I used some negative thin space}
   contracted with particular on-shell asymptotic states. For  $s=1$ vector  the  asymptotic states 
   are the standard helicity $\pm1$  states, while in the $s>1$ case  with the   free equation (in TT gauge)  
   $\del^{2s} h_s=0$ describing   total of 
   $s(s+1)$ dynamical degrees of freedom  one may choose a special solution 
   corresponding, {\em e.g.}, to the standard massless  
    helicity $\pm s$ field.\foot{For example, in the case of the 
    Weyl graviton one can always solve the linearised Bach equations by  imposing the linearised Einstein equations. 
   One may also consider  other special choices of  solutions of $\del^4$ equations as asymptotic states.} 
  % cf. \ci{Riegert:1984hf}.}

%New
Before turning to the discussion of higher spin  terms in \rf{2}   let us first recall the    structure of the non-linear 
terms for the low ($s=0,1,2,$) spins only. Instead of starting with the scalar  action  involving only the linear  coupling to the 
background fields $h_0,h_1,h_2$ (which here we assume to be subject to  TT condition and drop total derivatives)
  %, {\em i.e.}  
\be \la{3}
L= -\del_\m \vp^* \del^\m \vp + \sum_s  h_s  \vp^* \J_s \vp\te  =  \del_\m \vp^* \del^\m \vp 
+  h_0  \vp^* \vp  + {i} h^\m \vp^* \del_\m \vp    + \ha h^{\m\n}  \del_\m   \vp^* \del_\n \vp   + ... \ee
let us start with the standard  manifestly  ($U(1)$, reparametrisation  and  Weyl) covariant 
coupling of a complex scalar field  to the background 
metric $g_{\m\n} = \eta_{\m\n}  + h'_{\m\n}$, vector field $ h'_\m$ %=   h_\m +...$  
 and a scalar $ h'_0$, {\em i.e.}
\be \la{4}  
I= \int d^4 x\te  \sqrt g \Big[- g^{\m\n} D_\m \vp^* D_\n  \vp  + (h_0'-{1\ov 6} R) \vp^* \vp\Big] \ , \ \ \ \qquad 
 D_\m  \vp = \del_\m \vp +  { i\ov 2} h'_\m \vp \ . \ee
The log UV divergent  part of the resulting scalar determinant (cf. \rf{1}) 
 is given by the standard covariant 
Seeley  coefficient expression 
%\ci{'tHooft:1974bx}
(we ignore unimportant overall  constant related to coupling $\kg$ in \rf{2})\foot{Note that this  action may be interpreted as the bosonic sector of $\mc N=1$ conformal supergravity action with $h_0'$ playing the role of the auxiliary field.}
 %  but keep track of relative 
\be \la{5}
S[h_0', h'_1,h'_2] =  \int d^4 x \te  \sqrt g \Big( h_0'^2  - {1 \ov 24}  F'^2_{\m\n}   + { 1\ov 60}   C^2_{\m\n\l\r} \Big)\ ,  \ee
where $F'_{\m\n}=\del_\m h'_\n - \del_\n h'_\m   $ and $C$ is the Weyl tensor for $g_{\m\n}$. 
Since the fields $h'_s$ in \rf{4}  are related  to $h_s$  in \rf{3} by a   local   non-linear  redefinition
 \ba\la{6}\te 
&\te  h_0'= h_0 + {1\ov 4}  h_\m h^\m +   {1\ov 96} (  \del_\l h_{\m\n} \del^\l h^{\m\n}  + 2h_{\m\n} \Box h^{\m\n}  
+ 2  \del_\l h_{\m\n} \del^\m h^{\l\n}  ) +...\ , 
 %+  {1\ov 16} ( h_{\m\n} \Box h^{\m\n}  + \del_\l h_{\m\n} \del^\l h^{\m\n} ) + ... \ , \ 
 \\
 &\no 
   \te   h'_\m= h_\m + {1\ov 2} h_{\m\n} h^\n +  % {1\ov 2} 
{1\ov 4}    h_{\m\n} h^{\n\l}  h_\l + ...   \ , \qquad 
 \   \ \ \   h'_{\m\n} = {1\ov 2}  h_{\m\n}  +  {1\ov 4}  h_{\m\l} h^\l_\n  % + ...
  - {1\ov 16} \eta_{\m\n} h^{\l\r} h_{\l\r}   + ...  \ , \end{align}
  expanding \rf{5}   we may  thus read off 
  the cubic  and quartic   couplings of the original $h_0,h_\m,h_{\m\n}$ fields
 in \rf{2}.    %(see Appendix \ref{A})

  In particular,   using \rf{6} %that  $h_0'= h_0 + {1\ov 4}  h_\m h^\m  + ...$
 we find that  the scalar-vector   sector  of \rf{5}  takes the  form 
 \be \la{7}
S[h_0, h_1] =  \int d^4 x \te   \Big[  ( h_0 + {1\ov 4} h_\m h^\m)^2  -  {1 \ov 24}  F_{\m\n}^2   \Big]   \ . \ee
 Thus the  simplest cubic and quartic vertices are $011$ and $1111$. 
 We also  conclude, in particular, that  the  contribution of the $h_0$ exchange to  the 
 4-vector scattering amplitude 
 cancels against  the 4-vector contact vertex. As there is no 3-vector coupling, the 
 full 4-vector tree-level amplitude should thus  be  given by the sum of all 
 exchanges of CHS fields  with $s\geq 2$  and happens to vanish as we will find in  section 3. 
  Similarly, the  112    vertex is  related  to the one in the Maxwell-Weyl  theory, the 
   222  and 2222   vertices  are related to those  in the Weyl theory, etc.  
 %  combined with  scalar $h_0$ exchange   contributes the same as 2222  vertex in Weyl theory. 
 Thus the  contribution to the 4-graviton  amplitude  computed 
 from the $s=0,2$ exchanges and the 2222   vertex should   be
   the same as the 4-graviton  amplitude in pure Weyl theory
 (that happens to vanish). The contributions of all $s>2$ CHS exchanges vanishes  separately as we 
 shall demonstrate in section 5. 
 
%New

This paper is organised as follows. In section 2 we  shall present  the results for 
some  cubic  CHS vertices  (relevant  for  the computation of   spin 1 and spin 2 scattering amplitudes below)
 from the  UV singular parts of the  scalar loop integrals,  with some  details relegated to  Appendix \ref{A}. 
In section 3 we shall compute the 4-vector scattering amplitude  and demonstrate that 
after summing  over all CHS exchanges  it  vanishes. 
We shall then  observe in section 4  that the conformal higher spin  exchange  amplitudes should have the  same  general structure as the partial wave amplitudes in  the representation of \ci{Jacob:1959at}.
In section 5   we shall find  that the   scattering amplitudes 22 $\to $ 22   and 11$\to $ 22 
 involving conformal gravitons  do have this  expected structure  and they also vanish once one  sums up all intermediate CHS exchanges. 
 Some concluding remarks   will be made in section 6. 
 In Appendix \ref{B} we shall independently   verify the vanishing of the 11$\to $ 11    amplitude  at special 
 kinematics (backward  scattering)   and find that this vanishing appears to generalise to the case of jj $\to$ jj 
 scattering with all ${\rm j} =1,2,3,...$  supporting  our 
 conjecture that  the   full  4-particle  S-matrix in CHS theory should be trivial. 
 In Appendix \ref{C} we shall give the general derivation of the expression for  the 
  CHS spin $s$  exchange contribution to the 11 $\to $ 11   amplitude.

%%%%%%%%%%%%%%%%%%%%%%%%%%
\section{Vertices  in  induced   conformal higher spin  action}
%%%%%%%%%%%%%%%%%%%%%%%%%%%%

To be able to compute the CHS scattering amplitudes we   should 
first determine  the relevant  cubic  and quartic terms in the "induced" action \rf{2}. 
We shall use the following notation  for totally symmetric  tensors: $J_{\m(s)} \equiv J_{\m_1...\m_s}$  and also 
$\del_{\m(s)} \equiv \del_{\m_1}...\del_{\m_s}$.
Our  starting point will be the  complex scalar  Lagrangian in external CHS background (see, {\em e.g.},  \ci{Bekaert:2010ky,Joung:2015eny}
and refs. there) 
%The current $J_{\mu(s)}(\varphi, \varphi^{*})$ is given by the general formula
\begin{align}
\mathscr L &= -\partial_{\mu}\varphi^{*}\,\partial^{\mu}\varphi+
\sum_{s=0}^{\infty} %\te \frac{1}{s!}\,
J_{\mu(s)}\,h^{\mu(s)} \ , \la{2.1}\\
J_{\mu(s)}(x) &={\te  \frac{i^{s}\,2^{s}\,s!}{(2s)!}\,}\sum_{k=0}^{s}\te 
\binom{s}{k}\binom{\frac{s+k-1}{2}}{s}\,G^{(k)}_{\mu(s)}\ , \la{2.2}\\
G^{(k)}_{\mu(s)} &= \Big[
(\partial-\partial')_{\mu(k)}(\partial+\partial')_{\mu(s-k)}\varphi(x)\,\varphi^{*}(x')
\Big]_{x=x'} \ ,  \la{2.3}
\end{align}
where   the low-spin   currents $J_{\mu(s)}$  are   
\be
\la{2.4}
\begin{split}
J &=\varphi\, \varphi^{*}, \qquad 
J_{\mu} =\te  \frac{i}{2}\,(\varphi^{*}\partial_{\mu}\varphi-\varphi\,\partial_{\mu}\varphi^{*}), \\
J_{\mu\nu} &=\te  \frac{1}{6}\,\big[\partial_{\mu}\varphi\,\partial_{\nu}\varphi^{*}
+\partial_{\nu}\varphi\,\partial_{\mu}\varphi^{*}  - \ha ( \varphi^{*} \partial_{\mu}\del_{\nu}\varphi\,+\varphi\,\partial_{\mu}\del_{\nu}
\varphi^{*})    \big]\ . 
\end{split}
\ee
%New 
 The   vertices in  the CHS action \rf{2}   may be  thought of as originating from  the   coinciding-point  limits
 of the current correlators in the  free scalar CFT \  $\langle J_{s_1}(x_1)  ...J_{s_n}(x_n)  \rangle\big|_{x_i \to x}$
 and can be  found in coordinate space  using, e.g., differential  regularisation  \ci{Freedman:1991tk}.
 %,Freedman:1992gr}.
 Here 
we shall   use  momentum space representation and 
dimensional regularisation ($d=4-\varepsilon$)  and  define the  classical CHS action $S[h]$  as in \rf{2}, {\em i.e.} 
as the UV pole part of the one-loop scalar $\vp$ effective action:
\be \la{02}
\Gamma[h]=  \frac{1}{(4\,\pi)^{2}\ \varepsilon}\,\SS[h]+
\text{finite} \ . 
\ee
In general, the CHS   action should contain  an arbitrary   dimensionless constant $\kg$ as in \rf{2}  that
 will  then appear  as $\kg^2$  factor in the resulting 4-particle  tree-level amplitude; in what   follows we shall ignore this universal overall 
 factor, {\em i.e.} set $\kg=1$. 
 
 We shall also assume that   the    background fields $h_s \equiv h_{\mu(s)}$  in \rf{2.1} are 
 transverse and traceless (TT) as this will be sufficient   for the subsequent 
 computation of the  on-shell scattering amplitudes.\foot{In 
 %v2
 contrast to usual massless  Fronsdal HS fields (where one can only fix  transverse or de Donder gauge off shell)  for the 
   conformal higher spin  fields the gauge  symmetry involves   both  the differential  and  the algebraic symmetry allowing one to fix TT gauge, and this  leads to substantial simplifications.}
%  \comment\red{I wonder whether we have to comment that for our CHS 
% fields there is  no double trace condition one has with Fronsdal fields}
 Note that in this case  we may  integrate by parts   to write the interaction terms as 
   $h^\m J_\m  \to    i h^\m \varphi^{*}\partial_{\mu}\varphi, 
    \ \    h^{\m\n} J_{\m\n} \to     {1\ov 2} h^{\m\n}  \del_\m \varphi^{*}\del_{\nu}\varphi  $ or $-{1\ov 2} h^{\m\n}   \varphi^{*} \del_\m \del_{\nu}\varphi $, etc. 
 In general, for TT fields 
  the $h_s (p'-p) \vp^*(-p')  \vp(p)$   vertex  in momentum representation   reduces simply to
\be \la{2.5}
V_{\mu(s)} (p) =\te  \frac{1}{s!}\, p_{\mu_{1}}\cdots p_{\mu_{s}} \ ,  \ee
where  $p$ is the momentum of the $\vp$ leg. 
We can then compute the  UV singular part of the  scalar loop  diagram with two $V_{\mu(s)} $ insertions
\be\la{f1}
 \vcenter{\hbox{\includegraphics[scale=0.45]{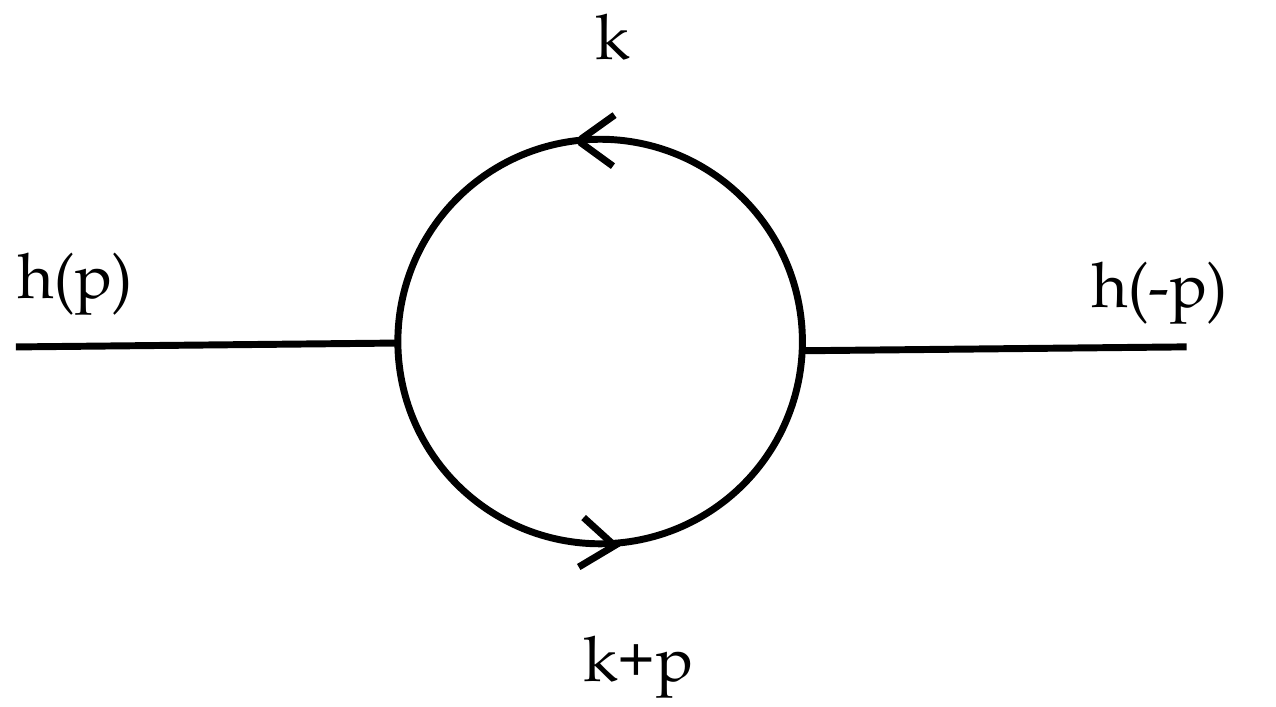}}}
%=\int \frac{d^dk}{(2\pi)^d}\ \frac{N(k,p)_{\dots}}{k^{2}\,(k+p)^{2}}
\ee
We then find that the  kinetic term in the CHS action (restricted to  TT fields) is   given by\foot{The relative normalisations  of the first $s=0,1,2$  terms here  are the same as in \rf{5},\rf{6}:
 note that $h'_{\m\n}=g_{\m\n} -\eta_{\m\n} $ in the  manifestly  covariant action \rf{4}  is  given by $h'_{\m\n}= {1\ov 2} h_{\m\n} + ...$
 in terms of $h_{\m\n}$ in \rf{2.1}   so that $C^2_{\m\n\l\r} \to  2 R^2_{\m\n} + ... \to  {1\ov 2} h'^{\m\n} \Box^2 h'_{\m\n}  \to {1\ov 8} h ^{\m\n} \Box^2 h_{\m\n}$.}
\be
\la{2.6}
\SS_2[h]= \sum_{s=0,1,2,...} \te 
\frac{1}{2^{s}\,(2s+1)!}\,\int d^{4}x\,
h_{\mu(s)}\,\Box^{s}\,h^{\mu(s)} \ . 
\ee
To  determine  the cubic $h_{s_1} h_{s_2} h_{s_3} $   couplings in the CHS action \rf{2} we are to 
 compute the  UV singular part of the one-loop scalar  diagram with  three (spin $s_1$, $s_2$ and $s_3$)   current vertex \rf{2.5}  insertions
\be
\la{f2}
\vcenter{\hbox{\includegraphics[scale=0.5]{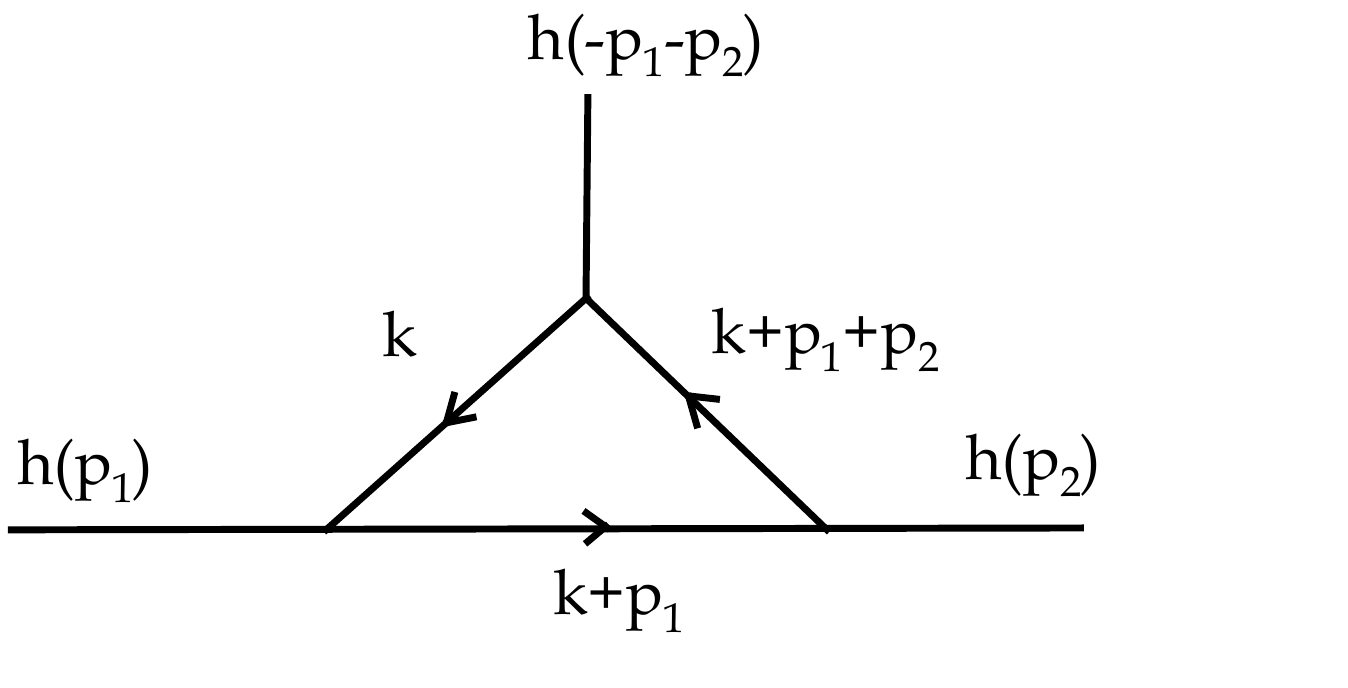}}}
\ee
As each spin $s$ vertex involves $s$ derivatives, parity invariance  implies that the resulting interaction 
  is non-zero only if 
$s_1+s_2+s_3 =$  even. 
One   can  also show  (using that $h_s$  is subject to the  TT  condition and  dimensional analysis)
that the 0-0-$s$  interaction   vanishes   for all $s$,  % (the corresponding  diagram \rf{f2}  has no UV  pole part), 
{\em i.e.} $\SS_3 [h_0, h_0, h_s] =0$.

For $s_1=s_2=1$  the  interaction  1-1-$s$   is non-zero only if  $s$  is even.
Written in coordinate space  the corresponding   cubic interaction  in the CHS action \rf{2}  (restricted again to TT fields) 
is found to be (see Appendix \ref{A})\foot{The last two terms  involving $\Box h^{\mu}$, {\em i.e.}  proportional to 
the vector field equation of motion  can be, in principle, 
 redefined away.} 
\begin{align}
& \SS_3 [h_1, h_1, h_s] ={\te \frac{(-1)^{s/2}}{(s+2)!}}\int d^{4}x\Big[
\partial_{\rho(s)}h_{\mu}h^{\mu}h^{\rho(s)}
-2 h_{\mu}\, \partial^{\mu}\,\partial_{\rho(s-1)}\,h_{\nu}\, h^{\nu\rho(s-1)}\no \\
&\te  - \partial_{\lambda}\partial^{\rho(s-2)}h^{\mu}\partial^{\lambda}h^{\nu}h_{\mu\nu\rho(s-2)} 
-\frac{s}{2} \partial^{\rho(s-2)}\Box h^{\mu}h^{\nu}h_{\mu\nu\rho(s-2)}
-\frac{s}{2} \partial^{\rho(s-2)}h^{\mu}\Box h^{\nu}h_{\mu\nu\rho(s-2)}
\Big]\la{2.66}\ . 
\end{align}
This vertex has total of $s$ derivatives in agreement with the  general structure of the CHS action  \rf{2}. 
In particular,     %for spin $s=0$, we recover (\ref{2.26}) that reads in coordinate space 
\be\la{2.7}
 \SS_3 [h_1, h_1, h_0] ={\te \frac{1}{2}}\int d^{4}x\,
 h_{\mu}h^{\mu}h_0 \ 
 \ee
  is in agreement   with \rf{5},\rf{7}. %     where  $h_0'=h_0 + {1\ov 4} h_\m h^\m$. 
  We  may  also  compute   the 4-vector quartic vertex from the  UV pole part of the diagram 
\be\la{f3} 
\vcenter{\hbox{\includegraphics[scale=0.25]{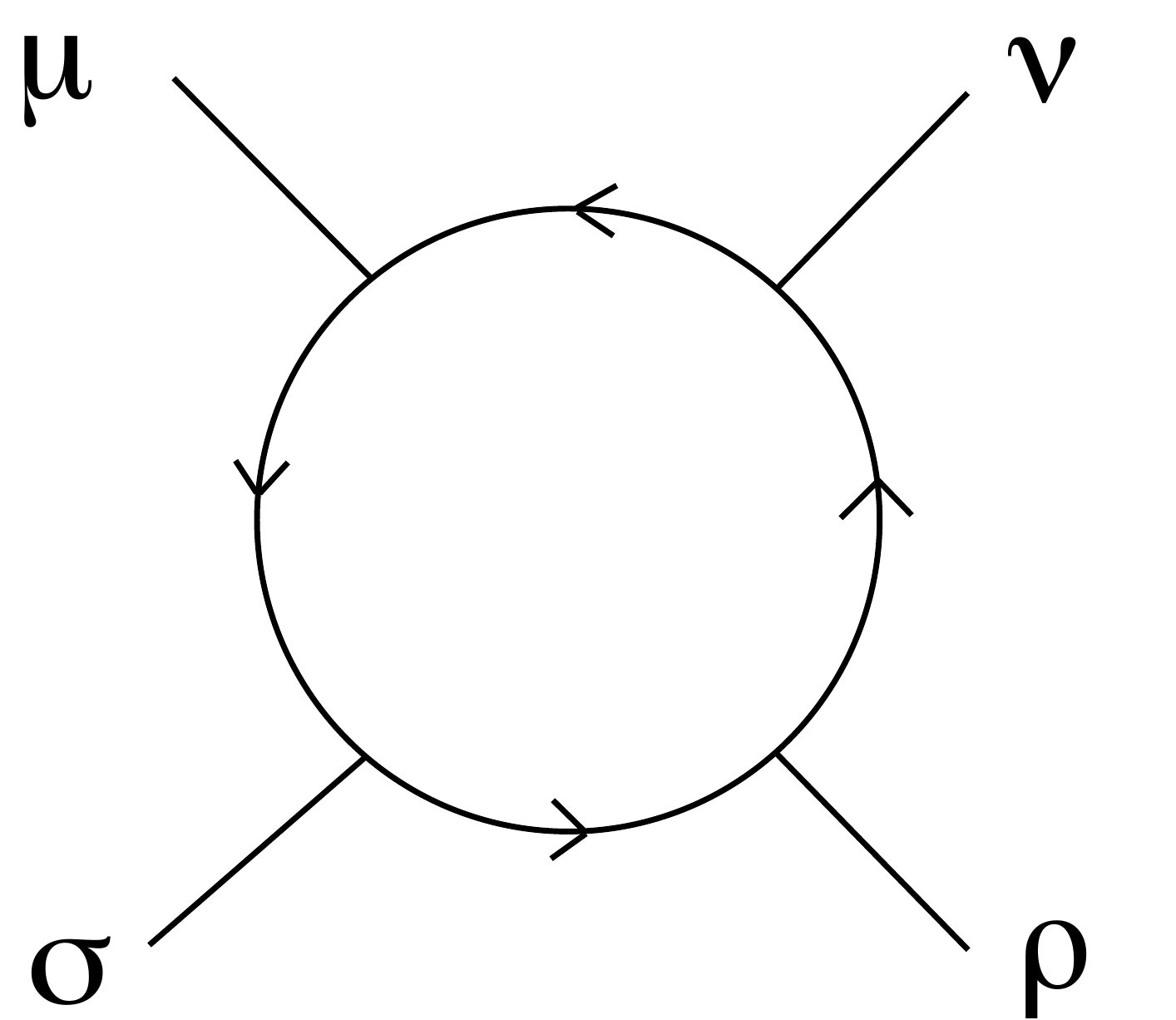}}}
\ee
getting, in agreement with \rf{7}, 
\be\la{2.99}
\SS_4 [h_1, h_1, h_1,h_1] 
  = {\te \frac{1}{16}}\,\int d^{4}x (h_{\mu}h^{\mu})^{2} \ .
\ee
   The vector-vector-graviton  coupling in \rf{2.66}  
\be\la{2.8} 
 \SS_3 [h_1, h_1, h_2] = {\te  \frac{1}{24}} 
 \int d^{4}x\Big[
\partial_{\rho}h_{\mu}\,\partial_{\sigma}h^{\mu}h^{\rho\sigma}
-2 \partial_{\rho}h_{\mu}\, \partial^{\mu}\,h_{\nu}\, h^{\nu\rho} +  \partial_{\r} h^{\mu}\partial^{\r}h^{\nu}h_{\mu\nu} 
 + 2 h^{\mu}\,\Box h^{\nu}h_{\mu\nu}
\Big]
\ee
 is equivalent (for TT fields) to  the  standard  graviton-Maxwell   coupling  in \rf{5}
 provided  one takes into account  the redefinitions in \rf{6}.   
% (taking  again into  the rescaling of gravitons in \rf{6}). 

Similar  expressions   are found  when the vector vertices in \rf{f2}  are replaced by the graviton ones, {\em i.e.} 
    for the   case of the $s_1=s_2=2, \ s_3=s$ interaction term  (see Appendix \ref{A}). 
With  $p_1,p_2$  being   spin 2   momenta   the resulting   2-2-$s$  interaction vertex  contains $s+2$ powers of momentum
and  reads 
%in momentum space is found to be
$
\text{V}_{\mu_{1}\mu_{2}, \nu_{1}\nu_{2}, \rho(s)} = V_{\mu_{1}\mu_{2}, \nu_{1}\nu_{2}, \rho(s)}
(p_{1}, p_{2})+V_{\mu_{1}\mu_{2}, \nu_{1}\nu_{2}, \rho(s)}
(p_{2}, p_{1}),
$
where\foot{Here we  drop terms 
proportional to equations   of motion for spin 2 states  as we will be using this vertex to compute 2-2-2-2  scattering   amplitude.
%\comment
Note, however, 
 that  for  spin 2  exchange one  is to use the 2-2-2  vertex that is symmetric in the three spin 2 legs with no on-shell condition assumed. % as since this is not valid for the internal particle.
% }} 
} %({\bf with the spin 2 legs on-shell})
\be \begin{split}\la{2.10}
&V_{\mu_{1}\mu_{2}, \nu_{1}\nu_{2}, \rho(s)}
(p_{1}, p_{2}) = {\te  \frac{1}{8\,(s+4)!}\,}
\Big[
- \sum_{\mu\neq\mu',\nu\neq\nu'} \eta_{\mu\nu}\,p_{2\mu'}\,p_{1\nu'}\,(p_{1})_{\rho(s)}\\
&+2\,\sum_{\mu\neq \mu'}\,p_{2\mu'}\,p_{1\nu_{1}}\,p_{1\nu_{2}}\,\eta_{\mu\rho_{1}}p_{1\rho_{2}}\dots p_{1\rho_{s}}
-2\,\sum_{\nu\neq \nu'} p_{1\nu'}\,p_{2\mu_{1}}\,p_{2\mu_{2}}\,\eta_{\nu\rho_{1}}p_{1\rho_{2}}\dots p_{1\rho_{s}}\Big]\\
&{\te -\frac{p_{1}\cdot p_{2}}{16\,(s+4)!}} \,\Big\{ 2(
\eta_{\mu_{1}\nu_{1}}\eta_{\mu_{2}\nu_{2}}+ \eta_{\mu_{1}\nu_{2}}\eta_{\mu_{2}\nu_{1}} )\,(p_{1})_{\rho(s)}\\ &- 4  
(\,p_{1\nu_{1}}\eta_{\mu_{1}\nu_{2}}\eta_{\mu_{2}\rho_{1}}   - p_{2\mu_{1}} \eta_{\mu_{2}\nu_{1}}\eta_{\nu_{2}\rho_{1}} + \sym\ \mu_{1,2},\nu_{1,2} ) \,p_{1\rho_{2}}\cdots p_{1\rho_{s}}
\\
& + \Big[ 6\,(\eta_{\mu_{1}\rho_{1}}\eta_{\mu_{2}\rho_{2}}+\eta_{\mu_{1}\rho_{2}}\eta_{\mu_{2}\rho_{1}})\,
p_{1\nu_{1}}\,p_{1\nu_{2}}
+6\,
(\eta_{\nu_{1}\rho_{1}}\eta_{\nu_{2}\rho_{2}}+\eta_{\nu_{1}\rho_{2}}\eta_{\nu_{2}\rho_{1}})\,
p_{2\mu_{1}}\,p_{2\mu_{2}}\\
&- \sum_{\mu\neq\mu',\nu\neq\nu'}
4\,
(\eta_{\mu\rho_{1}}\eta_{\nu\rho_{2}}+\eta_{\mu\rho_{2}}\eta_{\nu\rho_{1}})\,
p_{2\mu'}\,p_{1\nu'}\Big]\,p_{1\rho_{3}}\cdots p_{1\rho_{s}}\Big\}\\
&
+ {\te \frac{(p_{1}\cdot p_{2})^{2}}{8\,(s+4)!} }  \Big\{
\sum_{\mu\neq\mu', \nu\neq\nu'}
(\eta_{\mu\rho_{1}}\eta_{\nu\rho_{2}}\eta_{\mu'\nu'}+\eta_{\mu\rho_{2}}\eta_{\nu\rho_{1}}\eta_{\mu'\nu'})\,p_{1\rho_{3}}\cdots p_{1\rho_{s}} 
\\
& - (
p_{1\nu_{1}}\, \eta_{\mu_{1}\rho_{1}}\eta_{\mu_{2}\rho_{2}}\eta_{\nu_{2}\rho_{3}}
-p_{2\mu_{1}}\,\eta_{\mu_{2}\rho_{1}}\eta_{\nu_{1}\rho_{2}}\eta_{\nu_{2}\rho_{3}}  + \sym\  \rho_{1,2,3} )
\,p_{1\rho_{4}}\cdots p_{1\rho_{s}}\Big\}\\
&-{\te \frac{(p_{1}\cdot p_{2})^{3}}{32\,(s+4)!}\,}
 (\eta_{\mu_{1}\rho_{1}}\eta_{\mu_{2}\rho_{2}}\eta_{\nu_{1}\rho_{3}}\eta_{\nu_{2}\rho_{4}}+
\sym\ \rho_{1,2,3,4})\,\,p_{1\rho_{5}}\cdots p_{1\rho_{s}}\ , 
\end{split}
\ee
where $\sym$   stands for adding terms ensuring symmetry in $(\m_1,\m_2)$, $(\nu_1,\nu_2)$   and $(\r_1, ...,\r_s)$. 
In particular, choosing $s=0$ we find   that  the 2-2-0  coupling term in the CHS action can be written as 
\be \la{2.111}
S_3[h_0,h_2,h_2] = {  \te {1\ov 48} }  \int d^4x\  h_0 \big( \del_\r h_{\m\n} \del^\r h^{\m\n}    + 2 \del_\r  h_{\m\n} \del^\m h^{\r\n}\big) 
\ . 
 \ee
One can  trace the origin of  this term to $h'^2_0$ term in \rf{5}   and the   redefinition \rf{6}
(in particular, it corresponds to  cross-term $h_0 R$  with $R$  in \rf{4} expanded to quadratic order in $h'_{\m\n}$).

The 1-0-$s$  vertex  multiplying $h_{\mu}(p_{1}), \ h_0(p_{2})$ and $  h_{\rho(s)}(-p_{1}-p_{2})$
is non-zero when $s$ is odd    and is  found to be (where   symmetrisation in $\r_i$ is assumed) 
\be
\la{2.11}
\begin{split}
\text{V}_{\mu,\rho(s)} &=\te  \frac{2}{(s+1)!}\,
\eta_{\mu\rho_{1}}\,p_{\rho_{2}}
\cdots p_{\rho_s} \ . 
\end{split}
\ee
Here $p$ stands  for either $p_1$ or $p_2$ (\rf{2.11}  is   symmetric 
under $p_{1}\to p_{2}$ as  the fields are assumed to be  TT and  $s$ is odd).
Similarly, the 2-0-$s$  vertex (non-vanishing for  $s$=even) is  given by 
\be\la{2.12}
\begin{split}
\text{V}_{\mu_{1}\mu_{2},\rho(s)}\te = &\te \frac{1}{(s+2)!}\,
\Big[-
(\eta_{\mu_{1}\rho_{1}}\,p_{1\, \mu_{2}}+\eta_{\mu_{2}\rho_{1}}p_{1\, \mu_{1}})\,p_{1\, \rho_{2}}
... p_{1\, \rho_{s}}\\ &\te \qquad \quad 
-\ha {p_{1}\cdot p_{2}}\,(
\eta_{\mu_{1}\rho_{1}}\eta_{\mu_{2}\rho_{2}}+\eta_{\mu_{2}\rho_{2}}\eta_{\mu_{1}\rho_{1}})\,
p_{1\, \rho_{3}}... p_{1\, \rho_{s}}
\Big]\ . 
\end{split}
\ee
In the case of 1-2-$s$
 vertex  (with $s$=odd)   appearing   multiplied   by the  TT fields 
$h_{\mu_{1}\mu_{2}}(p_{1}),$ $ h_{\nu}(p_{2}),$ $ h_{\rho(s)}(-p_{1}-p_{2})$
we get 
%${\rm V}_{\mu_{1}\mu_{2},\nu, \rho(s)} (p_1,p_2) = V_{\mu_{1}\mu_{2},\nu, \rho(s)} (p_1,p_2) 
%+ V_{\mu_{1}\mu_{2},\nu, \rho(s)} (p_2,p_1)$ where 
\be
\la{2.13}
\begin{split}
&\V_{\mu_{1}\mu_{2},\nu,\rho(s)}(p_{1},p_{2}) =\te \frac{1}{(s+3)!}\Big\{
(\eta_{\mu_{1}\nu}p_{2\mu_{2}}+\eta_{\mu_{2}\nu}p_{2\mu_{1}}) p_{1\rho(s)}\\
&+(-\eta_{\mu_{1}\rho_{1}}p_{2\mu_{2}}p_{1\nu}
-\eta_{\mu_{2}\rho_{1}}p_{2\mu_{1}}p_{1\nu}
+2\eta_{\nu\rho_{1}}p_{2\mu_{1}}p_{2\mu_{2}})\,p_{1\rho_{2}}... p_{1\rho_{s}}\\
&-(p_{1}\cdot p_{2})\Big[
(\eta_{\mu_{1}\nu}\eta_{\mu_{2}\rho_{1}}+\eta_{\mu_{1}\rho_{1}}\eta_{\mu_{2}\nu})\,p_{1\rho_{2}}... p_{1\rho_{s}}\\
&+\Big( (\eta_{\mu_{1}\rho_{1}}\eta_{\nu\rho_{2}}+\eta_{\mu_{1}\rho_{2}}\eta_{\nu\rho_{1}})\,p_{2\mu_{2}}
 -(\eta_{\mu_{1}\rho_{1}}\eta_{\mu_{2}\rho_{2}}+ \eta_{\mu_{2}\rho_{1}}\eta_{\mu_{1}\rho_{2}})\,
p_{1\nu}
\\
&+ (\eta_{\mu_{2}\rho_{1}}\eta_{\nu\rho_{2}}+\eta_{\mu_{2}\rho_{2}}\eta_{\nu\rho_{1}})
\,p_{2\mu_{1}}\Big)\,p_{1\rho_{3}}... p_{1\rho_{s}}\Big]\\
&\te +\frac{1}{3}(p_{1}\cdot p_{2})^{2}
(\eta_{\mu_{1}\rho_{1}}\eta_{\mu_{2}\rho_{2}}\eta_{\nu\rho_{3}}+\sym \ 
\rho_{1,2,3})\,p_{1\rho_{3}}... p_{1\rho_{s}}\Big\} \ . 
\end{split}
\ee

%%%%%%%%%%%%%%%%%%%%%%%%%%%
\section{Scattering   %for vectors and gravitons 
 in CHS theory:     4-vector  amplitude}
%%%%%%%%%%%%%%%%%%%%%%%%%%%%%

We can now use the   interaction terms in the CHS   action found in the previous section to compute some  tree-level 
scattering amplitudes. As the scalar $h_0$  is non-propagating, {\em i.e.} has  zero on-shell value we will not discuss
analogs of scattering amplitudes with $h_0$ on external legs. 

The vector $h_1$   has the standard Maxwell  kinetic term, so 
the definition  of the  corresponding 4-vector scattering amplitude is standard (the  same as in the case of the 
external scalar scattering  in \ci{Joung:2015eny}):  we consider physical helicity $\pm 1$   photon states 
on external lines  and include  all  exchanges  with two 1-1-s   vertices \rf{2.6}   connected  by 
TT propagator for even-spin $s$  CHS  field. The contribution of the $h_0$ exchange  
 due to 011 vertex \rf{2.7} 
exactly  cancels  against the contact 4-vector vertex \rf{2.99} as follows from \rf{7}
so it remains to consider only  the exchanges with $s=2,4,6,...$    fields on internal lines. 

Before proceeding with spin 1 scattering   let  us note for the future  discussion in sections 4 and 5 
 that as  the CHS fields  with $s>1$  in  \rf{2}  have higher-derivative kinetic terms, 
 the notion  of S-matrix   for $s>1$  external lines requires  special definition. 
 Given  the free  spin $s>1 $ CHS equation in TT gauge  $\Box^s  h_s=0 $  one can always 
 choose a special  solution $h_s=h^{(0)}_s $  satisfying $\Box  h^{(0)}_s=0 $.
 %\foot{For example, the  Bach equations of the Weyl theory 
 %are  solved  by all solutions of  the Einstein equations, both at linear and non-linear level.} 
 This equation  has  further on-shell gauge  invariance allowing one  to reduce the number of independent  solutions 
   to just  2   of a standard  2-derivative massless  particle.  In what  follows 
 we shall always consider  only these special "physical"  helicity $\pm s$ 
 modes as the asymptotic states  in the definition of 
 the CHS S-matrix.\foot{Ideally, 
 %New 0707.4437
 one would  like to  start with a formulation of the CHS  theory  in terms of the set of fields 
 with ordinary (2-derivative) kinetic terms that exists at the  quadratic  level  \ci{Metsaev:2007fq,Metsaev:2007rw}.   
 Unfortunately,  an  existence of  such local  action at the interacting level   is  an open
 question   for $s>2$. % (and seems unlikely). 
 }
 Thus the asymptotic states will always  be massless on-shell particles with 
  %be the standard massless  particles with  two $\pm s$  helicity states and 
 $p^2=0$  while the internal  spin $s'$  
 propagator will be ${1\ov p^{2s'}}$ times the  TT projector.\foot{The condition $p^2=0$   for the  external lines 
 will help
  to simplify  the expressions for the
 required  cubic CHS vertices.}

 %%%%%%%%%%%%%%%%%%%
 \subsection{4-vector  exchange  amplitude}
 
Let us start with the 4-vector scattering amplitude  and first   set up  the notation we will use. 
We shall   consider  the scattering process 
\be\la{fig4}
\vcenter{\hbox{\includegraphics[scale=0.2]{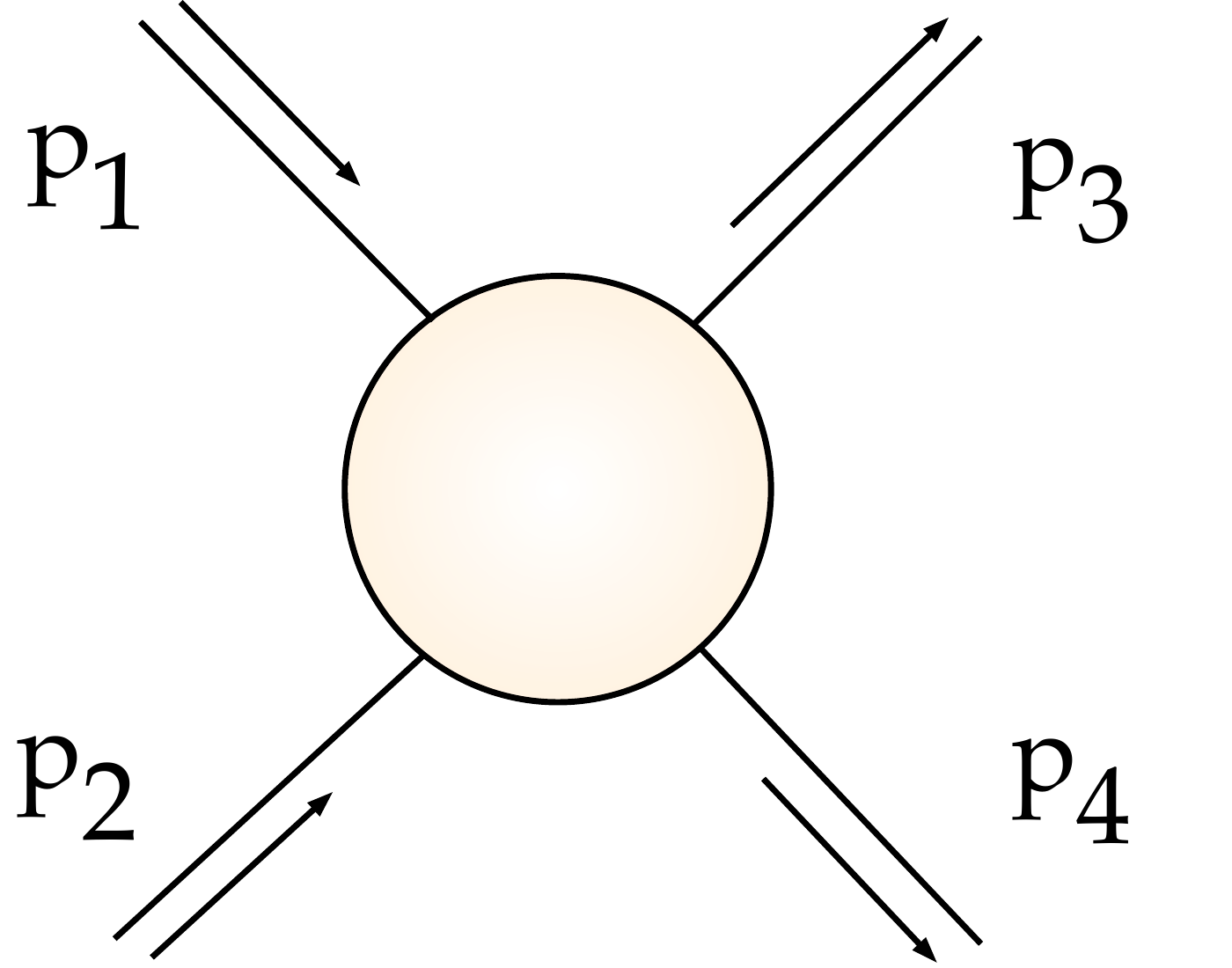}}}
=  \vcenter{\hbox{\includegraphics[scale=0.4]{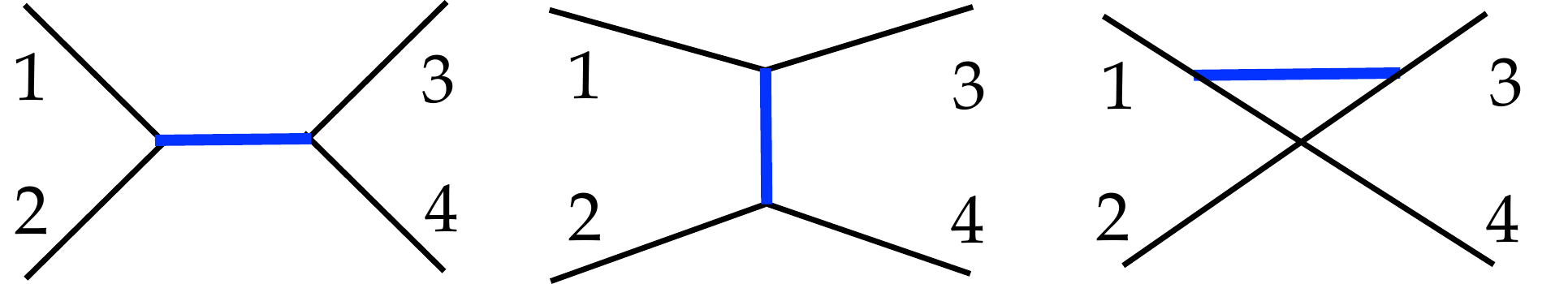}}}
\ee
with 
 $(\lambda_{1},p_1), ( \lambda_{2},p_2)\to (\lambda_{3},p_3),  (\lambda_{4},p_4)$, {\em i.e.} assume two momenta and helicities as incoming and two as outcoming with $\l_i=\pm 1$ and $p^2_i=0$. 
In the c.o.m.  frame, we have  for the  momenta and polarisation vectors\footnote{The helicity $\pm 1$ polarisation vector for an { initial} state with 
$p=(\omega, \omega\sin\theta, 0, \omega
\cos\theta)$ is $\varepsilon^{(\pm)}_{\mu}(p)=\mp{1 \ov \sqrt 2} (0,\cos\theta,\pm i, -\sin\theta)$.
If the state is {final}, the polarisation vector is $(\varepsilon^{(\pm)}_{\mu}(p))^{*}$
(see, {\em e.g.},  \cite{Elvang:2013cua,Gleisberg:2003ue}).
} 
\be\la{3.1} 
\begin{aligned}
p_{1} &= (\omega, 0, 0, \omega), && 
& \varepsilon_{1}(p_{1}) &= -\sqt \lambda_{1}\,(0,1,i\,\lambda_{1},0) \\
p_{2} &= (\omega, 0, 0, -\omega), && 
&\varepsilon_{2}(p_{2}) &= -\sqt \lambda_{2}\,(0,-1,i\,\lambda_{2},0) \\
p_{3} &= (\omega, \omega\,\sin\theta, 0, \omega\,\cos\theta), && 
&[\varepsilon_{3}(p_{3})]^{*} &= -\sqt \lambda_{3}\,(0,\cos\theta,-i\,\lambda_{3},-\sin\theta) \\
p_{4} &= (\omega, -\omega\,\sin\theta, 0, -\omega\,\cos\theta), && 
&[\varepsilon_{4}(p_{4})]^{*} &= -\sqt \lambda_{4}\,(0,-\cos\theta,-i\,\lambda_{4},\sin\theta) \\
\end{aligned}
\ee
and  the Mandelstam variables are 
\ba
&\qquad  \rs =-(p_{1}+p_{2})^{2} = 4\omega^{2}, \qquad\qquad
\rt =-(p_{1}-p_{3})^{2} = -2\omega^{2}\,(1-\cos\theta),\no  \\ &
 \qquad
\ru =-(p_{1}-p_{4})^{2} = -2\omega^{2}\,(1+\cos\theta)\ , \qquad \qquad  \rs+\rt + \ru =0 \ .  \la{3.2} 
\end{align}
The exchange diagrams involve two 11s  vertices  corresponding to   $h_1(p)\, h_1(q)\,  h_s(-p-q)$   from \rf{2.6}\foot{Here we use  that 
$p^2=q^2=0$ for the  external vector lines; $p=p_1, \, q=p_2$ in s-channel, etc.}
\be
\la{3.3}
\begin{split}
&\text{V}_{\alpha,\beta,\rho(s)}(p, q) =\te  \frac{1}{(s+2)!}\Big\{
\eta_{\alpha\beta}\big[\tfrac{1}{2}\,(p)_{\rho(s)}+\tfrac{1}{2}\,(q)_{\rho(s)}\big]\\
&-\tfrac{1}{2}\eta_{\alpha\rho_{1}}p_{\beta}p_{\rho_{2}}\dots p_{\rho_{s}}
+\tfrac{1}{2}\eta_{\beta\rho_{1}}q_{\alpha}p_{\rho_{2}}\dots p_{\rho_{s}}
-\tfrac{1}{2}\eta_{\beta\rho_{1}}q_{\alpha}q_{\rho_{2}}\dots q_{\rho_{s}}
+\tfrac{1}{2}\eta_{\alpha\rho_{1}}p_{\beta}q_{\rho_{2}}\dots q_{\rho_{s}}\\
&-\tfrac{1}{2}\,\eta_{\alpha\rho_{1}}\eta_{\beta\rho_{2}}
\,p_{\rho_{3}}\dots p_{\rho_{s}}\, p\cdot q
-\tfrac{1}{2}\eta_{\alpha\rho_{1}}\eta_{\beta\rho_{2}}
\,q_{\rho_{3}}\dots q_{\rho_{s}}\,p\cdot q
\Big\}.
\end{split}
\ee
Here   $h_s$ is assumed to be in TT   gauge   with the corresponding propagator (cf. \rf{2.6}) 
\be\te \la{3.4} 
D^{\alpha(s)}_{\beta(s) }(p) = \frac{2^{s-1}(2s+1)!}{(p^{2})^{s}}\ \P^{\alpha_{1}\cdots \alpha_{s}}_{\beta_{1}\cdots \beta_{s}}(p),
\ee
where  the TT projector $\P^{\alpha(s)}_{\beta(s) }$ is built out of  products of 
$\P^{\alpha}_{\beta}= \delta^{\alpha}_{\beta}-\frac{p^{\alpha}\,p_{\beta}}{p^{2}}$, 
{\em e.g.,}\footnote{In general (cf. \rf{c3})   %in $d$ dimensions 
$$\Pi^{\n(s)}_{\m(s)} = \sum_{l=0}^{[\frac{s}{2}]} \aa_{s,l}\, \ M^{\n(s-2l)}_{\m(s-2l)}\, N^{\n(2l)}_{\m(2l)} \ , \ \ \ \qquad \qquad   
\aa_{s,l} = \frac{(-1)^l s! \, \G(s-l + {1\ov 2}   )}{2^{2l} (s-2l)! \, l! \,\G(s + {1 \ov 2} ) }
$$
where 
$M_{\m(p)}^{\n(p)} = {\Pi^{(\n_1}_{\m_1} \ldots \Pi^{\n_p)}_{\m_p}}$ and 
$
N_{\m(2q)}^{\n(2q)} =\Pi_{(\m_1 \m_2} \ldots \Pi_{\m_{q-1} \m_q)} \,   \Pi^{(\n_1 \n_2} \ldots \Pi^{\n_{q-1} \n_q)}
$.
}
\ba
\la{3.9}
%P^{\alpha}_{\beta}(p) &= \delta^{\alpha}_{\beta}-\frac{p^{\alpha}\,p_{\beta}}{p^{2}}, \\
\P^{\alpha_{1}\alpha_{2}}_{\beta_{1}\beta_{2}} &= \te \P^{\alpha_{1}}_{(\beta_{1}}\,\P^{\alpha_{2}}_{\beta_{2})}-\frac{1}{3}\,
\P^{\alpha_{1}\alpha_{2}}\P_{\beta_{1}\beta_{2}}, \\
\P^{\alpha_{1}\alpha_{2}\alpha_{3}\alpha_{4}}_{\beta_{1}\beta_{2}\beta_{3}\beta_{4}} &\te = 
\P^{(\alpha_{1}}_{(\beta_{1}}P^{\alpha_{2}}_{\beta_{2}}P^{\alpha_{3}}_{\beta_{3}}
\P^{\alpha_{4})}_{\beta_{4})}
 -\frac{6}{7}\,P^{(\alpha_{1}\alpha_{2}}\P_{(\beta_{1}\beta_{2}}\P^{\alpha_{3}}_{\beta_{3}}\P^{\alpha_{4})}_{\beta_{4})}+\frac{3}{35}\,
 \P^{(\alpha_{1}\alpha_{2}}\P^{\alpha_{3}\alpha_{4})}\P_{(\beta_{1}\beta_{2}}
\ \P_{\beta_{3}\beta_{4})}\ , \ \   {\rm etc.}\no 
 \end{align}
%For   general $s$  the TT  projector  is given by a similar  type of  products  of $P^{\alpha}_{\beta}$ with 
%  the coefficients   given by 
%\be\la{3.10} \te 
%\frac{\big(-\frac{1}{4}\big)^{\ell}\,s!\,\Gamma(-\ell+s+\frac{1}{2})}{\ell!(s-2\ell)!\,\Gamma(s+\frac{1}{2})}, \qquad\qquad  \ell=0, 1, \dots, \frac{s}{2}.
%\ee
The resulting  s, t, u-channel amplitudes are are given by\footnote{The factors of 2  in the vertices   are due to the 
symmetry of the  external  lines: for a Lagrangian term $\Phi^{n}=n!\frac{\Phi^{n}}{n!}$, the  standard 
Feynman rules imply the  coefficient $n!$.}
\be\la{3.10}
\begin{split}
A_\rs^{(s)} &= 2\,\V(p_{1}, p_{2})\cdot D(p_{1}+p_{2})\cdot 2\,\V(p_{3}, p_{4})\cdot \varepsilon_{1}\,
\varepsilon_{2}\,\varepsilon_{3}^{*}\,\varepsilon_{4}^{*},\\
A_\rt^{(s)} &= 2\,\V(p_{1}, p_{3})\cdot D(p_{1}-p_{3})\cdot 2\,\V(p_{2}, p_{4})\cdot \varepsilon_{1}\,
\varepsilon_{2}\,\varepsilon_{3}^{*}\,\varepsilon_{4}^{*},\\
A_\ru^{(s)} &= 2\,\V(p_{1}, p_{4})\cdot D(p_{1}-p_{4})\cdot 2\,\V(p_{2}, p_{3})\cdot \varepsilon_{1}\,
\varepsilon_{2}\,\varepsilon_{3}^{*}\,\varepsilon_{4}^{*} \ . 
\end{split}
\ee
Evaluating  these amplitudes for   various  helicity choices   we find  that all amplitudes where helicity is not conserved 
vanish\foot{In standard terminology that means that only MHV   amplitudes are non-zero. The same will be true for all amplitudes discussed below.} 
 and  for the helicity-conserving  cases $\pm\pm \to \pm \pm$  or the crossing-related cases $\pm \mp \to \pm \mp$ we get\footnote{The vanishing of
 the  s-channel exchange for the same helicity process  may be related to the fact that 
 % understood according to 
%the clue in \cite{Datta:2004mr}. For gravitational interaction,
 helicity is conserved in the 3-point 
vertices  where   one  has only the  $\pm\mp$  combination  (same happens for gravitational 
 interactions, cf. \cite{Datta:2004mr}).}
% Instead $\pm\pm$ can couple only with the
%trace of the metric. Similar should happen here with higher spin field -- we thank R. Roiban.
\be\la{3.11} 
\begin{split}
&\pm\pm\to  \pm\pm: \ \qquad A_{\rs}^{(s)}= 0, \qquad 
A_{\rt}^{(s)}= c_{s}\, \big(\frac{\rs}{\rt}\big)^{s}\,P_{s}\big(\frac{\rt}{\rs}\big), 
\qquad 
A_{\ru}^{(s)} = c_{s}\, \big(\frac{\rs}{\ru}\big)^{s}\,P_{s}\big(\frac{\ru}{\rs}\big), \\
&\pm\mp\to  \pm\mp: \ \qquad  A_{\rs}^{(s)}= c_{s}\, \big(\frac{\ru}{\rs}\big)^{s}\,P_{s}\big(\frac{\rs}{\ru}\big), 
\qquad 
A_{\rt}^{(s)} = c_{s}\, \big(\frac{\ru}{\rt}\big)^{s}\, P_{s}\big(\frac{\rt}{\ru}\big), \qquad
A_{\ru}^{(s)} = 0 \ . 
\end{split}
\ee
Like in the external   scalar scattering case \ci{Joung:2015eny}  the  scale invariance of the CHS theory 
and the fact that $h_1$   has canonical dimension 1  implies that the $d=4$ amplitude depends only on ratios of the 
Mandelstam variables. We have isolated  powers of  these ratios  containing  the  internal  spin $s$ CHS propagator factor in each channel
({\em i.e.} $ \rt^{-s} $ in t-channel,  etc.). 
The remaining momentum dependence   is given by the even  degree $s-2$ polynomials   $P_{s}(x)$ 
 whose normalisation   is fixed by the  condition $P_{2}=1$ and $  P_{s>2}(-1)=1$
\be
\la{3.12}
\begin{split}
P_{2}(x) &= 1, \qquad \qquad 
P_{4}(x) =28+42\,x+15\,x^{2}, \\
P_{6}(x) &= 495+1320\,x+1260\,x^{2}+504\,x^{3}+70\,x^{4}, \\
P_{8}(x) &= 8008 + 30030\, x + 45045\, x^2 + 34320\, x^3 + 13860\, x^4 + 2772\, x^5 + 
 210\, x^6, \\
P_{10}(x) &= 125970+604656 \,x+1225224 \,x^2+1361360 \,x^3+900900 \,x^4+360360 \,x^5\\
&\ \ \ \ \ +84084
   \,x^6+10296 \,x^7+495 \,x^8 \ . 
\end{split}
\ee
The overall numerical coefficients $c_s$  then are 
\be\la{3.13}\te 
c_{2} = \frac{5}{12}, \qquad 
c_{4} = \frac{1}{20}, \qquad 
c_{6} = \frac{13}{840}, \qquad 
c_{8} = \frac{17}{2520}, \qquad
c_{10} = \frac{7}{1980}.
\ee
The expressions \rf{3.12},\rf{3.13} were found  by direct computations for $s=2,...,10$  
but admit a  natural   generalisation  to any $s$.
The polynomials $P_{s}(x)$  may be expressed in terms of Jacobi polynomials $P_{n}^{(a,b)}(x)$
 as  ($s=2,4,6,\dots$)
\be
\la{3.14}\te 
P_{s}(x) = x^{s-2}\,P_{s-2}^{(4,0)}\big(\frac{x+2}{x}\big) \ , 
\ee
while  the simplest interpolating ansatz  for $c_s$ is 
\be \la{3.15}\te 
c_{s} = \frac{2\,(2\,s+1)}{(s-1)\,s\,(s+1)\,(s+2)} \ .
\ee 
We shall provide the general derivation  of \rf{3.14},\rf{3.15} in Appendix \ref{C}. 

Let us mention also  some useful alternative forms  of  the polynomials $P_{s}$  in \rf{3.14} 
\begin{align}
P_{s}(x) &\no = \sum_{j=0}^{s-2}\te (-1)^{j}\,\binom{s-2}{j}\,\binom{2s-j}{s+2}\,(1+x)^{s-2-j}\,x^{j}  = \binom{2s}{s+2}\,(1+x)^{s-2}\,{}_{2}F_{1}\big(2-s,2-s,-2s,\frac{x}{x+1}\big) \no \\
&{\te \ = \frac{x^{2\,s+1}}{(s-2)!}\,\big(\frac{d}{dx}\big)^{s-2}
\frac{(1+x)^{s-2}}{x^{s+3}}
=} \sum_{j=2}^{s}\te \frac{1}{(j-2)!\,(j+2)!}\,\frac{(s+j)!}{(s-j)!}\,x^{s-j} 
%(1+x)^{s-2}\,P_{s-2}^{(-2s-1,4)}\big(\frac{1-x}{1+x}\big) 
\la{3.155} 
\\
& \te 
 \no
 = \frac{1}{24}(s-1)\,s\,(s+1)\,(s+2)\,x^{s-2}\,{}_{2}F_{1}\big(2-s,s+3,5; -\frac{1}{x}\big).\notag
%\notag \\
%&= x^{s-2}\,P_{s-2}^{(4,0)}\Big(\frac{x+2}{x}\Big).	notag 
\end{align}
For comparison, in the case of the  external massless 
scalar scattering (coupled to CHS fields  as in \rf{2.1},\rf{2.2} or \rf{2.5}) 
  the s-channel $\varphi \varphi^* \to \varphi \varphi^*$ amplitude was     given \ci{Joung:2015eny} 
in terms of the Legendre polynomial ${\rm P}_s= P_{s}^{(0,0)}$ 
\be 
\la{555}
A^{(s)}_{\rs \,  \vp \vp^* \to \vp \vp^*} = (s+ \ha) \ {P_{s}^{(0,0)}} (-1-2x)  \ , \ \ \ \  \te \qquad   x={\rt\ov \rs} \ . \ee
One can also consider the "mixed"  scattering amplitude $\vp \vp^* \to 11$   of two external   conformal scalars into two  vectors 
due to exchange of the tower of CHS fields.\foot{Here we are assuming  that one  adds the action \rf{3}  of one  conformal scalar 
coupled to CHS fields  to the CHS action \rf{2}  and then studies the S-matrix of the resulting theory.}
In this    case the s-channel  even   spin s  exchange  amplitude  is given by (cf. \rf{3.10})
\be
A_{\rs \, \vp\vp^*\to 11}^{(s)} = \V_{\varphi\varphi^{*}s}(p_{1}, p_{2})\cdot D^{(s)}(p_{1}+p_{2})\cdot 2\,\V(p_{3}, p_{4})\cdot \varepsilon_{3}^{*}\,\varepsilon_{4}^{*}, \la{345}
\ee
where  $\V_{\varphi\varphi^{*}s}$ is the vertex in  \rf{2.5}    and  $\V$  is  the 11s  vertex  as in \rf{3.3},\rf{3.10}. 
The  resulting $\pm\pm$ amplitudes vanish while  the  helicity-preserving $\pm\mp$ ones  may be written as  (cf. \rf{3.11}) 
%The non zero amplitude may always be written
\be
\la{328}
A^{(s)}_{\rs \, \vp\vp^*\to \pm1\mp1} = k_{s}\,\frac{\rt\,\ru}{\rs^{2}}\,Q_{s}\big(\frac{\rt}{\rs}\big)\ ,  %,\qquad \qquad
%Q_{s}(1)=1.
\ee
where we ignore  the overall minus sign  and assume that  numerical coefficients  are defined by normalising the  order $s-2$ polynomial 
$Q_s$  as $Q_{s}(-1)=1$. Explicitly, one finds $Q_{2} = 1, \ 
Q_{4} = \frac{1}{3}\,(3+14\,x+14\,x^{2}), ...$ and $k_1= \frac{5}{2} , \ k_{4} = \frac{9}{2}, ...$.
On the basis of $s=2, ...,10$  examples can  guess   the general $s$ expressions as 
\be \la{329}\te 
Q_{s}(x) = \frac{2}{s\,(s-1)}\,P_{s-2}^{(2,2)}(-1-2\,x) \ , \ \ \ \ \qquad \ \ \qquad 
k_{s} =  s + \ha \ ,
\ee
where $P_{s-2}^{(2,2)}$ is again the Jacobi  polynomial  (cf. \rf{3.14},\rf{3.15},\rf{555}).

%%%%%%%%%%%%%%%%%%%%%%%%
\subsection{Summing over spins}

%New 
As we already mentioned above, the $s=0$ exchange  contribution cancels against 
 the one of the 1111  vertex \rf{2.99}.
Thus to get  the total   amplitude  it remains to sum 
over all spin $s=2,4, ...$ exchanges. 
%%%%%%%%%%%%%%%%%%%%%%%%
%\subsubsection{Sum over spin}
Let us  consider, {\em e.g.},  the $\pm\pm\to \pm\pm$   case in \rf{3.11}  (the discussion of the  
$\pm\mp\to \pm\mp$  case is similar)
% \comment\red{this should be instead $\pm\mp\pm\mp$ ok ? })
where the  sum over channels is 
\be
\la{3.16}
\pm\pm\to \pm\pm: \ \ \ \ \ \  A^{(s)} =  c_{s}\,\Big[\big(\frac{\rs}{\rt}\big)^{s}\,
P_{s}\Big(\frac{\rt}{\rs}\Big)
+\big(\frac{\rs}{\ru}\big)^{s}\,P_{s}\Big(\frac{\ru}{\rs}\Big)\Big].
\ee
Since $\ru=-\rs-\rt$   this may be written as a function of one variable  $x \equiv  {\rt\ov \rs} $ as 
\be\la{3.17} 
A^{(s)} = \sigma_{s}(x)+\sigma_{s}(-1-x), \qquad\qquad  \sigma_{s}(x) \equiv 
 c_{s}\,x^{-s}\,P_{s}(x)\ . \ee
 We may compute the sum over $s$  by introducing  first  an extra  regularisation parameter $z$  and defining 
 \be \la{3.18} 
 \sigma (x) \equiv 
  \lim_{z\to 1}  \sigma(x; z) \ , \ \ \ \ \ \qquad   \sigma(x; z) \equiv    \sum_{s=2,4,6,...}^{\infty} \sigma_{s}(x)\ z^{s-2}  \ . 
  \ee
Let us first  omit  the  overall coefficient $c_s$  in $\sigma_{s}$       and consider the formal sum 
over all (even and odd)   $s=2,3,4,...$
\be
K(x;z) \equiv \sum_{s=2}^{\infty} x^{-s}\,P_{s}(x)\,z^{s-2} \ . \la{3.19} 
\ee
This  can be  written in a closed form using the generating function for  the Jacobi polynomials $P_{s-2}^{(4,0)}$
\cite{koekoek1996askey} as 
\be
\la{3.22}\te 
K(x;z) = \frac{16}{x^{2}}\, \big[\sqrt{z^2-\frac{2 z (x+2)}{x}+1}\big]^{-1}  \big[\sqrt{z^2-\frac{2 z
   (x+2)}{x}+1}-z+1\big]^{-4}\ . 
\ee
Then using the fact that  $c_s$  in \rf{3.15}   admits the following representation 
\be\te 
c_{s} = \frac{1}{s+2}-\frac{1}{s+1}+\frac{1}{s-1}-\frac{1}{s} \ , \la{3.23}
\ee
we can  compute $\sigma(x; z)$   by multiplying \rf{3.22}  by  a suitable power of $z$, integrating,
and then dividing  by another  appropriate power of $z$. 
Finally, the sum over spins  may be restricted to even $s$ only  by 
simply taking one half  of the sum of the expressions with $z$ and  with $-z$. 
While the resulting expression  is quite 
cumbersome, its $z\to 1$  limit  turns out  to be   finite and  simple  
\be
\la{3.24}
\sigma(x) %&= \mathop{\sum_{s=2}^{\infty}}_{s\ \rm even}\sigma_{s}(x) = \lim_{z\to 1}\mathop{\sum_{s=2}}_{s\ \rm even}^{\infty}c_{s}\, x^{-s}\,P_{s}(x)\,z^{s-2} \\
=\te  x (x+1)\, \log \frac{x+1}{x} - x -\ha \ .
\ee
As  it is easy to check,  this function satisfies the relation $\sigma(x) = - \sigma(-1-x) $ 
implying that the total summed-over-spins   amplitude vanishes: 
\be\la{3.25} 
A (x)= {\sum_{s=2,4,6,...}^{\infty}} A^{(s)}(x) = \sigma(x)+\sigma(-1-x)=0 \ .
\ee
Here we formally assumed  that $\sigma(x)$   is defined    for any $x$   using  analytic continuation.
In fact, this  function is real for $x\in [-\infty,-1]\cup[0,\infty]$ while  the argument of the amplitude in  \rf{3.17}  is 
$
x ={ \rt\ov \rs}=-\frac{1}{2}\,(1-\cos\theta)\in[-1,0].
$
In the latter "physical"    interval one  finds again  that $A(x+i\,0)=0$ for any sign of the infinitesimal imaginary part.
%New
In Appendix \ref{B}   we provide an independent  check of the vanishing  of the  amplitude \rf{3.25} 
at the special   kinematical point $\ru=0$ or $x=-1$ (or, equivalently, at $x=0$). 

Another clarification  is  that  in the above discussion we have excluded  the special points $x=0,-1$
where   the amplitude may have delta-function   singularities   as in  the external scalar amplitude case \ci{Joung:2015eny}. 
Indeed, as was shown  in \ci{Joung:2015eny}, the   sum of the Legendre  polynomials  in \rf{555} 
 is given by $\sum_{s=0}^\infty (s+{1\ov 2} ) {\rm P}_s (x ) = \delta (x-1)$, so the total amplitude  given by  the  sum of the s- and t-channels 
 is $\sim  \delta({\ru\ov \rs} ) + \delta({\ru\ov \rt} ) $ which vanishes for  real momenta.
Similar  cancellation   happens here as well  as we show in  Appendix  \ref{B}.

%v2
%%%%%%%%%%%%%%%%%%%%%%%%%%%
\section{General structure of  CHS    exchange    amplitudes  }
%%%%%%%%%%%%%%%%%%%%%%%%%%%%%

To generalise  the above    vector scattering  results to  higher $s>1$   spin scattering case 
it is useful first to  discuss  the   structure of the  CHS 4-particle amplitudes expected on the basis 
of Lorentz and scale invariance.  It turns out that 
the appearance of the special Jacobi polynomials in \rf{3.14}, \rf{555}   and \rf{329} 
is  not accidental  and   may be related to the    partial wave expansion  of the 
$\l_1,\l_2 \to \l_3,\l_4$ transition   amplitude  discussed  by Jacob  and Wick  \cite{Jacob:1959at} (see also \cite{eden1967high,collins}).

Considering  the  c.o.m.  frame     and using the   completeness of states 
 relation one can represent  generic  scattering  amplitude as a  sum over  on-shell states  of a
massive particle   with  mass$=\sqrt \rs$   and spin $J$       \cite{Jacob:1959at}
\begin{align}\la{41}
&\quad A_{\{\lambda_i\} }(\rs, \theta) =  R_{\{\lambda_i\}} (\theta) 
\sum_{J\geq M}(J+\tfrac{1}{2})\,  \F^{(J)}_{\{\lambda_i\}}(\rs) 
\te 
\ P_{J-M}^{(|\lambda-\mu|, |\lambda+\mu|)}(\cos\theta)\ , 
%\end{split}
\\
&\qquad \la{42}
\lambda=\lambda_{1}-\lambda_{2},\ \  \ \mu=\lambda_{3}-\lambda_{4}, \ \ \ \qquad
M=\max(|\lambda|, |\mu|) \ , 
\\  \la{43}
&\quad  \te  R_{\{\lambda_i\}}(\theta)= \big(\cos\frac{\theta}{2}\big)^{|\lambda+\mu|}\,\big(\sin\frac{\theta}{2}\big)^{|\lambda-\mu|}= \big(-\frac{\ru}{\rs}\big)^{\frac{1}{2}\,|\lambda+\mu|}\,\big(-\frac{\rt}{\rs}\big)^{\frac{1}{2}\,|\lambda-\mu|} \ . 
\end{align}
Here  ${\{\lambda_i\}}=(\lambda_{1}, \lambda_{2}; \lambda_{3}, \lambda_{4})$,\ \  $\cos \theta = 1  + 2 {\rt\ov \rs}$   and $P^{(a,b)}_k$  is the Jacobi polynomial. 
The latter  originates   from the expression for   the spherical $d$-function  ($N=\min(|\lambda|, |\mu|) $)
\be d^J_{\l\m}(\theta) = \te  \sqrt{ (J+M)! (J-M)! \ov (J+N)! (J-N)! } \ R_{\{\lambda_i\}}(\theta) \ P_{J-M}^{(|\lambda-\mu|, |\lambda+\mu|)}(\cos\theta)
\la{433}\ . 
\ee
 We  assume    that the scattering particles are massless. %  and  omit the  trivial rotation factor $e^{i(\l-\m)\phi}$.
% (or  consider the  scattering plane with $\phi=0$).
  If the theory is scale-invariant  (has no dimensional parameters) 
 the dependence  of the coefficient functions  $\F^{(J)}_{\{\lambda_i\}}$  on $\rs$ should be controlled only by dimensions  $\Delta_i$ 
  of the scattering fields,  
\be \la{44} 
\F^{(J)}_{\{\lambda_i\}} (\rs) =  F^{(J)}_{\{\lambda_i\}}\, \rs ^\Delta \ , \ \ \ \ \ \ \ \qquad  \ \ \ \   \Delta\equiv  \ha (4-\sum^4_{i =1}\Delta_i) \ . \ee 
Here $F^{(J)}_{\{\lambda_i\}}$  are some numerical coefficients  encoding  dynamical  information 
about a  particular theory. %  under consideration. 
For example,  $\Delta=0$   for scattering of dimension 1  massless  scalars  or vectors in 4d.
In the case of  CHS  scattering with asymptotic states  chosen,  as discussed in the previous section,
 to be the  standard massless  spin $|\l_i|$  particles  the representation \rf{41},\rf{44}  should again apply  with 
  $\Delta= \ha (\sum_i |\l_i|- 4)$  (the CHS  field dimensions are $\Delta_i = 2 - |\l_i|$).   
  
  Our  key observation is that   in the present context of conformal higher spin   theory 
        the     spin  $J$  contribution 
  in \rf{41}  should have  the same structure as  %can be  formally interpreted 
    the contribution of  an intermediate CHS   field  exchange with $s=J$ in s-channel. 
  This should be a kinematical consequence of the fact that 
   a   massive ($m^2=\rs$)   intermediate spin $J$   state in \rf{41}   may be described  by a   totally symmetric 
  field $\phi_{\m_1...\m_J}$ satisfying $(-\Box + m^2)\phi_{\m_1...\m_J}=0$ as well as the   tracelessness and transversality 
  conditions   (leaving  only $2J+1$  states as physical  degrees of freedom).
   At the same   time,  the  CHS  scattering is  also  mediated   by the   TT  field exchange  with 
   the propagator in \rf{3.4}. The only  formal  difference is in the  overall   $\rs$-dependence  that appears in  $\F$  
    but in the CHS scattering case  the latter   is controlled   by the  scale invariance leading to \rf{44}.
    
    %New
   % It   should be noted that this c
   This   formal  interpretation  of the spin $J$ term in \rf{41}  as the   CHS  spin $s=J$ 
   exchange  amplitude should directly apply only to the s-channel  exchange: this is due to the selection 
   of $\rs$  variable as the c.o.m.  frame  mass parameter  in \rf{41} and thus as 
   the variable  that should appear  in the propagator of the  corresponding exchanged CHS      field. 
  %   Thus  the total CHS amplitude given by the sum over channels 
  % as in \rf{fig4},\rf{3.16} and then   summing over all   spins    will not directly have the form \rf{41}, {\em i.e.}  to put it in the form \rf{41} 
%   will require some resummation. \comment\red{I would not say resummation and also would separate the problem of multiple channels from that of summing over spin. I would say : 
The total CHS amplitude given by the sum over all channels    as in \rf{fig4},\rf{3.16} will also have the general form  \rf{41}  
when expanded  in the  Jacobi polynomials  but the $J=s$ identification of the  particular term in the sum 
with the   contribution of the CHS exchange   will be  valid only in a 
  particular  channel (in s-channel or after renaming the kinematic variables and helicities also in t- and u-channels, see below).

    %Besides, after summing over the internal exchanged spin, the simple
  % organisation  \rf{41} of the separate exchanges will be even more hidden.

    Another remark is  that  this  identification of the   $J$-term    in \rf{41} with  the higher spin exchange  does not apply 
    to the case of  the 2-derivative massless   higher spin scattering in flat space discussed in \ci{Ponomarev:2016jqk}. The reason is that 
     the massless  spin $s$  propagator (taken, {\em e.g.}, in the de Donder  gauge) is not  traceless-transverse
    and thus   the massless higher spin particle exchange  cannot be directly  identified  with a massive spin $J$  on-shell  state 
    contribution in the sum in \rf{41}.\foot{Indeed, the scattering 
    amplitude for  four massless   scalars  exchanging the tower of massless  higher spins 
      was given in  \ci{Ponomarev:2016jqk} by the sum of the Chebyshev polynomials  rather than the 
      Legendre polynomials 
    appearing \rf{555} in the case  of  the conformal higher spin exchange in \ci{Joung:2015eny}.
    Interestingly, there is  still a formal relation  between  4-scalar scattering via massless  higher spin exchange in $d+1$ dimensions  and 
   via conformal higher spin exchange  in $d$ dimensions   suggesting  possible  AdS/CFT  connection.
   }

Let us now   see how the   previously discussed   cases of the external   scalar and vector scattering \rf{3.11},\rf{555},\rf{328}
 via the CHS exchange  may be related to  \rf{41}. 
In the case of the  $\varphi\varphi^{*}\to \varphi\varphi^{*}$ scattering  we have 
$
\lambda_{i}=0, \ \lambda = \mu = 0, \ M=0,
$
and thus  should expect, according to \rf{41},\rf{44}, to find the s-channel spin $J$  contribution to be 
 %(identifying $J=s$)
\be
\la{4.8}
\begin{split}
A^{(J)} _{\rs\, 0,0; 0,0}(\rs, \cos\theta) = 
%\sum_{J=0}^\infty
 (J+\tfrac{1}{2}) F^{(J)}_{0,0; 0,0} \,P_{J}^{(0,0)}(\cos\theta)\ .
\end{split}
\ee
Comparing  this with the s-channel result \rf{555} of the direct computation   
using that $\cos\theta=
1+2\,\frac{\rt}{\rs} $
we  conclude  that % the s-channel 
%(and also u-channel)  
the two expression indeed match  provided  $s=J$ and 
\be \la{4411}
F^{(s)}_{0,0; 0,0}=1 \ . \ee
For  $\varphi\varphi^{*}\to {\pm}{\mp}$   in  \rf{345},\rf{328}  we have 
$
\l=\lambda_{1}=\lambda_{2}=0, \ 
\lambda_{3}=-\lambda_{4} = \pm 1, \  \mu = M= \pm 2,
$
and thus  from  \rf{41},\rf{44}  should get 
\be
\la{411}
\begin{split}
A^{(J)}_{\rs\, 0,0; \pm 1,\mp 1}(\rs, \cos\theta) &= 
(J+\tfrac{1}{2}) F^{(J)}_{0,0; \pm 1 , \mp 1} \,\frac{\ru\,\rt}{\rs^{2}}
\,P_{J-2}^{(2,2)}(\cos\theta) \ . 
\end{split}
\ee
%Comparing with (\ref{3.28}) and (\ref{3.31}) , we find that only even $s$ are exchanged and 
%\footnote{Notice that for even $s$ the polynomial $P_{s-2}^{(2,2)}(x)$ is even.}
Comparing this with \rf{328},\rf{329}  we find   perfect match    
 provided  $s$ is identified with $J$ (which should be  taken to be  even)  and\foot{Note
 that % Also,
  the restriction that $J=s$ should be 
 even  does not follow from \rf{41}   and is an extra dynamical  property of CHS theory (parity invariance of the original 
 scalar theory \rf{2.1} implying the absence 1-1-$s$  vertices with odd $s$).
 For even $s$ the polynomial $P_{s-2}^{(2,2)}(x)$ is even.}
\be\la{412}
F^{(s)}_{0,0;\pm  \mp } = \frac{2}{s\,(s-1)}.
\ee
%%%%%%%%%%%%%%%%%%%%%%%%%
In the case of  ${\pm}1 {\pm}1\to {\pm}1 {\pm}1 $   scattering in \rf{3.11} 
we have the  two  contributions of the t- and u- channels 
 that  are  to be analysed separately. 
      For example, considering  the t-channel exchange  to be able to compare it  to \rf{41} 
      we should  first re-interpret it as an s-channel exchange   by relabelling the states and Mandelstam variables. 
      Explicitly,   the t-channel   scattering   of original "X"-particles may be  interpreted as s-channel  scattering of 
effective  "Y"-particles, {\em i.e.} 
$
X_{1}+\overline X_{3}\to X_{4}+\overline X_{2}  $ is equivalent to 
$Y_{1}+Y_{2}\to Y_{3}+Y_{4}.
$
For the Y-particles  we then  have 
$
\lambda_{1}=-\lambda_{2}=\pm 1, \ 
\lambda_{3}=-\lambda_{4} = \pm 1, \ 
\lambda = \mu = 2,  \ M=2
$
and thus  from \rf{41},\rf{44} we should get 
%The predicted angular dependence is -- in terms of kinematics o fthe Y particles -- 
\be\la{47}
\begin{split}
A^{(J)}_{\rs \, \pm 1, \mp 1; \mp 1, \pm 1}&= 
 (J+\tfrac{1}{2}) F^{(J)}_{\pm 1, \mp 1; \mp 1, \pm 1} \,\frac{\ru_{Y}^{2}}{\rs_{Y}^{2}}
\,P_{J-2}^{(0, 4)}(\cos\theta_{Y}),\qquad \cos\theta_Y = -1-2\,\frac{\ru_{Y}}{\rs_{Y}}\ .
\end{split}
\ee
The Y-kinematics becomes the X-kinematics after 
$\rs_{Y}\to \rt$, $\rt_{Y}\to \ru$, $\ru_{Y}\to \rs$. Thus  for the t-channel exchange 
of the X-particles  we should  get   %have 
\be
\la{4.16}
\begin{split}
 A^{(J)}_{\rt  \, \pm 1, \pm 1; \pm 1, \pm 1}(\rt, \cos\theta)&= 
(J+\tfrac{1}{2}) F^{(J)}_{\pm 1, \pm 1; \pm 1, \pm 1} \,\frac{\rs^{2}}{\rt^{2}}
\,P_{J-2}^{(0, 4)}(-1-2\,\frac{\rs}{\rt})\ .
\end{split}
\ee
This matches the t-channel result  in \rf{3.11},\rf{3.14},\rf{3.15}   with $J=s$   since 
%On the other hand, the t-channel is from (\ref{3.19}) proportional to 
\be\la{49}
\big(\frac{\rs}{\rt}\big)^{s}\,\big(\frac{\rt}{\rs}\big)^{s-2}\,P_{s-2}^{(4,0)}\big(1+2\,\frac{\rs}{\rt}\big)
%= \frac{\rs^{2}}{\rt^{2}}\,P_{s-2}^{(4,0)}\big(1+2\,\frac{\rs}{\rt}\big) 
= 
\frac{\rs^{2}}{\rt^{2}}\,P_{s-2}^{(0,4)}\big(-1-2\,\frac{\rs}{\rt}\big) \ , 
\ee
%that agrees with (\ref{4.16}) upon identification $J=s$. The Jacob-Wick amplitude is finally
provided also we  choose 
\be\la{410}
F^{(s)}_{\pm 1, \pm 1; \pm 1, \pm 1} =  {c_s \ov s+ {1\ov 2}} = \frac{4}{(s-1)\,s\,(s+1)\,(s+2)} \ .
\ee
%The matching   in the u-channel can be demonstrated similarly.
%\section{A conjecture for the Jacob-Wick coefficient}
Guided by the above three  examples  \rf{4411},\rf{412}  and \rf{410}  we may   conjecture the
 general  dependence of $F^{(J)}_{\{\lambda_i\}}$ in \rf{44} 
on $J=s$  in the case of CHS exchange amplitudes  to be%  (cf. \rf{433})
\foot{One reason   why this choice may be special 
is the following property of the Jacobi polynomials: 

\noindent
$
(\lambda+\mu)!\,\frac{(s-M)!}{(s+N)!}\,P_{s-M}^{(|\lambda-\mu|, \lambda+\mu)}(-1) = 1
$ where $\l$ and $\m$ should be integer and $s-M$ should be even integer.
%\comment\red{
This implies  the $s$-independence of the backward scattering amplitude in 
the direct channel.}
\be
\la{5.1}
F^{(s)}_{\{\lambda_i\}} = \rc_{\lambda,\mu}\,\frac{(s-M)!}{(s+N)!}, \qquad\qquad  N =\min(|\lambda|,|\mu|),\ M=\max(|\lambda|, |\mu|) \ . 
\ee
%where $\mc N_{\lambda, \mu}$ does not depend on $s$. 
Then  for $ \rc_{\lambda,\mu}=1$ we   indeed  get 
$F^{(s)}_{\{0\}}$  as in \rf{3.14}, $F^{(s)}_{0,0; \pm\mp}$ as  in \rf{412} 
  and $F^{(s)}_{\{\pm 1\}}$ as  in  \rf{410} (where one should, as explained above,   use $\l=\m=2$ 
to match the t-channel result). 
%New
 It turns out that 
  this ansatz \rf{5.1}  applies also in all other cases   discussed  below.

%%%%%%%%%%%%%%%%%%%%%%%%

%%%%%%%%%%%%%%%%

\section{Scattering amplitudes   with conformal gravitons }  
%and  spin 2  scattering
%%%%%%%%%%%%%%%%%%%%%%%%%

Let us  now turn to the discussion   of  conformal graviton scattering due to the exchange of the CHS   fields. 
The relevant 2-2-$s$ interaction vertex was given in \rf{2.10}. 
As discussed in section 3, we shall be scattering only the  "physical" massless  spin 2  component 
of  the conformal spin 2  field, attaching the corresponding asymptotic  states to the 
amputated Green's functions.\foot{Let us note also   that 
%v2
  using the standard    formulation  of the CHS  action \rf{2} one may also  study
  the scattering of other  "ghost"-like modes described by the higher-derivative
 CHS equations. 
 In general, 
the  
6  dynamical degrees of freedom of the Weyl graviton  way be 
described  by the  collection of  the standard  massless spin 2 field,  massless vector, and   massless 
spin 2 ghost field  states 
 (for  a  discussion of  solutions of linearised Weyl gravity equations  reproducing the 
 dynamical   degrees of freedom 
 count \ci{Fradkin:1981iu,Fradkin:1981jc,Lee:1982cp}  see   \ci{Riegert:1984hf} and also Appendix C in \ci{Metsaev:2007fq}).
%One may, in principle also study the scattering of spin 1 and ghost spin 2   components of the Weyl graviton. 
% (see \cite{Riegert:1984hf}). 
Explicitly,  choosing the   TT  gauge    ($h_m^m=0, \ \del^mh_{mn} =0$) 
 we get the free conformal graviton  equation as 
 $\Box^2  h_{mn}=0 $    which is  solved by   $ h_{mn} = h^{(1)}_{mn} +h^{(2)}_{mn}    
   = (a_{mn} + b_{mn}  u_k x^k) e^{ip \cdot x}  + c.c. $  where 
 $p^2=0  , \      u^2=-1  ,  \  u \cdot p \not=0\ , \ \ \   a^m_m = b^m_m =0   $.
 Here 
 $h^{(1)}_{mn}$  represents the  massless  spin 2 and spin 1 modes  and 
  $h^{(2)}_{mn} $  the ghost-like   spin 2   mode (which grows in time  and leads to   negative energy contributions). 
 Using the Lorentz symmetry  and the  residual gauge freedom one may  choose   \ci{Riegert:1984hf}: \ 
 $p^m=(p,0,0,p) , \ \   u^m=(1,0,0,0) \ $, \ 
 $  a_{11}+a_{22}=b_{11}+b_{22}=0\ , \ \ \  a_{m3}=b_{m3}=b_{m0}=0  $ 
 and then 
 the   2+2+2    dynamical  d.o.f.   are described   by the   helicity $\pm 2$ tensor 
   $(a_{11} \pm i a_{12} ) e^{ip \cdot x}$,  %:  \ \ \ \ \  \ \   \ \  \ physical $\l=\pm 2$   massless  tensor  
   helicity $\pm 1$  vector 
  $(a_{01} \pm i a_{02} ) e^{ip \cdot x}$  %:  \ \ \ \ \  \ \ \ \  \ physical $\l=\pm 1$   massless vector 
 and  helicity $\pm 2$     ghost   tensor
  $(b_{11} \pm i b_{12} )  x^0 e^{ip \cdot x}$. %:  \ \ \ \ \  \ \   ghost  \ \ \ \  \ $\l=\pm 2$   massless    tensor  
  The spin 1 and ghost spin 2   become parts of massive spin 2 ghost 
  if one adds the $R$ term to Weyl action to get a diagonal mode decomposition. 
 At  the level  of the flat-space partition   function of Weyl graviton the above  2+2+2   
 split corresponds to   the following 
decomposition \ci{Fradkin:1985am}:
$$Z_2= \Big[
\frac{(\det\Delta_{1})^{3} }{(\det \Delta_{2})^2}
\Big]^{1/2}
=  \Big[
\frac{\det\Delta_{1\, \perp}}{\det\Delta_{2\, \perp}}
\Big]^{1/2}  \Big[
\frac{\det\Delta_{0\, \perp}}{\det\Delta_{2\, \perp}}
\Big]^{1/2}    = \Big[
\frac{\det\Delta_{1\, \perp}}{\det\Delta_{2\, \perp}}
\Big]^{1/2}   \Big[
\frac{\det\Delta_{0\, \perp}}{\det\Delta_{1\, \perp}}
\Big]^{1/2}     \Big[
\frac{\det\Delta_{1\, \perp}}{\det\Delta_{2\, \perp}}
\Big]^{1/2} 
$$
Here $\Delta_s$ are 2-derivative Laplacians defined on traceless  rank  $s$ symmetric fields.
}
%%%%%%%%%%%%%%%%%%%%%%%%%%%%%%%%%%%%%

\subsection{$22\to 22$  scattering} 

Let us   first   discuss    what we should expect  to get for the  structure of the 22$\to$22     even spin $s\ge 4 $ exchange
on  general  symmetry grounds.  %We shall  consider separately   the cases of  $s\geq 4$  and $s=0,2$. 
We shall assume that as  in the 4-vector case the  non-vanishing  scattering  amplitudes   should  be similar to 
\rf{3.11} (where now $\pm$ will stand for $\pm 2$  helicities of the external massless   graviton state). Thus 
for the $++\to ++$  amplitude 
we should  have   the contributions from the t- and u- channels.\foot{We again assume  two   incoming and two outgoing momenta; 
choosing all momenta as incoming  this   becomes  the   MHV $++--$ amplitude.}
Then   repeating  the analysis that  in the vector case lead to \rf{4.16}
we conclude     that for the t-channel exchange  of  CHS spin $J$ states 
we should expect  from \rf{41},\rf{44} to find for $ J\geq 4$ 
%\item Internal exchange is restricted to {\bf even} spin $s\ge 4$. This must be due to a combination of parity
%constraints and cancellation with low spin contact terms or special 222 vertices.
%\end{enumerate}
%With these assumptions (to be checked later), 
\be
\la{6.2}
\begin{split}
A^{(J)}_{\rt \, \pm 2, \pm 2; \pm 2, \pm 2}(\rt, \cos\theta) = 
(J+\tfrac{1}{2}) F^{(J)}_{\pm 2, \pm 2; \pm 2, \pm 2}\,\rt^{2} \,\frac{\rs^{4}}{\rt^{4}}
\,P_{J-4}^{(0, 8)}(-1-2\,\frac{\rs}{\rt})\ .
\end{split}
\ee
Here $\rt^2$  factor reflects the   fact that the conformal   graviton has dimension 0  (cf. \rf{44}). 
%We can thus repeat the spin sum we did in the 1111 case
%where (\ref{3.16}) and (\ref{3.19}) now read \footnote{We do not optimise notation, just mimick the 1111 case.}
The total amplitude due to   spin $s=J$  exchange should then  be  as in \rf{3.16} (cf. \rf{3.14}) 
\begin{align}\la{6.3}
\pm2\pm2 \to \pm2\pm2: \qquad   A^{(s)} & = c_{s}\,\rs^{2}\,\big[\big(\frac{\rs}{\rt}\big)^{s-2}\,
 P_{s}\big(\frac{\rt}{\rs}\big)
+\big(\frac{\rs}{\ru}\big)^{s-2}\,P_{s}\big(\frac{\ru}{\rs}\big)\big], \\
& \ \ \ P_{s}(x) = \te x^{s-2}\,P_{s-4}^{(8,0)}\big(\frac{x+2}{x}\big)\ . \la{63}
\end{align}
If we also  assume the validity  of the conjecture \rf{5.1}  for the coefficients $F^{(J)}_{\{\l_i\}}$
 then we may expect also  to  get  
\be\la{64}\te
c_{s} =\rc\ \frac{2s+1}{(s-3)(s-2)(s-1)s(s+1)(s+2)(s+3)(s+4)} \  , 
\ee
where $\rc$ is  some  $s$-independent numerical factor.

Remarkably, the direct  computation  based on  the CHS  action and   carried out for several 
even\foot{Recall that the 2-2$s$ vertex  \rf{2.10}   vanishes  for odd $s$.} 
  values of $s\geq 4 $ 
confirms  the above expressions \rf{6.3},\rf{64}  and fixes  $\rc$ in \rf{64} to be 
\be \te 
\rc = {9\ov 8}   \ . \la{65} \ee  
Similar result is  found for  the $\pm2\mp 2\to \pm2\mp2$ exchange  (cf. \rf{3.11}).
The general derivation of \rf{6.3}--\rf{65} may be  given using the same  formalism as described  for the 11 $\to$ 11  case in Appendix \ref{C}.

We can now sum the amplitude \rf{6.3} over all  even $s=4,6,...$   using the same method as in  the vector scattering case
\rf{3.17}--\rf{3.24}:
\ba \la{66}
&\sum_{s=4,6,...}^\infty    A^{(s)} (x) = \rs^2\big[  \s(x)   + \s(-1-x) \big] \ , \ \ \ \ \  \qquad  x\equiv \frac{\rt}{\rs}\ ,   \\
& \s(x) =  \lim_{z\to 1}  \sum_{s=4,6,...}^\infty c_{s} \, x^{2-s}\,P_{s}(x)\,z^{s-4} \ . \la{67}
\end{align}
 After a rather involved computation  using the generating  function  for the Jacobi polynomials in \rf{63} 
 we found  that\footnote{Let us  note  a  similarity in the structure  of (\ref{66})  and  (\ref{3.24}). This suggests that for 
higher  spin  jj$\to$  jj  scattering one may  be able to guess 
the expression for $\sigma(x)$ and then check that the  coefficients in its expansion  in a suitable set of
Jacobi polynomials  reproduces the $c_{s}$  prefactor. Similar ideas have been 
exploited in \cite{gustavsson2001some}.}
\be
\la{6.6}
\sigma(x) = \te \frac{1}{4320}\Big[ 60\, (x+1)^3\, x^3\, \log \frac{x+1}{x}
-60\, x^5-150\, x^4-110\, x^3-15\, x^2+3\,
   x-1\Big] \ .
\ee
%Since \be \sigma(-\tfrac{1}{2}+y) = \frac{15 \big(4 y^2-1\big)^3 \log\big(\frac{2 y+1}{2 y-1}\big)-4 y \big(80
%   \big(3 y^2-2\big) y^2+33\big)}{69120} = -\sigma(-\tfrac{1}{2}-y), \ee
%we obtain 
One can then check that  the combination of  the 
$\s$ functions  appearing in \rf{66}  vanishes as in the vector exchange case \rf{3.25} 
\be\la{677}
\sigma(x)+\sigma(-1-x) = 0 \ , 
\ee
{\em i.e.}   the t- and u-channel   contributions  summed over $s=4,6,...$ 
cancel   against each other. 

To  find the total  $22\to 22$   amplitude 
%%%%%%%%%%%%%%%%%%%%%%%%%%%%%%
one is still to  add    (i)  the contributions of the low-spin $s <4$   CHS exchanges
({\em i.e.}  the exchange mediated by the non-propagating spin 0  field  $h_0$
and the   exchange  of the spin 2 conformal  graviton itself) 
and  also (ii) the  contribution of the  2222  contact vertex
that is found from the UV singular part of the diagram \rf{f3} with four  spin 2 current insertions
(with vertices in \rf{2.5} as the  external legs are assumed to be TT). 
We found the following expressions for the spin 0 exchanges   with the cubic vertex in \rf{2.111}: 
\be\la{667}
\begin{split}
&  \pm2 \pm2 \to  \pm2\pm2: \ \ \qquad  \te A_{\rs}^{(0)} = \frac{\rs^{2}}{4608}, \qquad
A_{\rt}^{(0)} = \frac{\rt^{2}\,\ru^{4}}{512\,\rs^{4}}, \qquad
A_{\ru}^{(0)} = \frac{\rt^{4}\,\ru^{2}}{512\,\rs^{4}}, \\
& \pm2 \mp2 \to  \pm2\mp2: \ \ \qquad   \te A_{\rs}^{(0)} = 0, \qquad\quad
A_{\rt}^{(0)} = \frac{\rt^{2}\,\ru^{4}}{512\,\rs^{4}}, \qquad
A_{\ru}^{(0)} = \frac{(\rs+3\rt)^{2}\,\ru^{4}}{4608\,\rs^{4}} \ . 
\end{split}
\ee
The spin 2 exchanges (with the 2-2-2  vertices  as in  \rf{2.10})
%being the same as in the Weyl theory) 
are  % found to be 
\ba \no 
&  \pm2 \pm2 \to  \pm2\pm2: \ \ \qquad  \te A_{\rs }^{(2)} = \frac{\rs^2+6\, \rs\, \rt+6\, \rt^2}{23040}, \quad
A_{\rt }^{(2)} =\frac{\ru^2 (2\, \rs^4-10\, \rs^3\, \rt
+33\, \rs^2\, \rt^2-24\, \rs\, \rt^3+
3\, \rt^4)}{7680\, \rs^4}\ , \\
&\qquad \qquad \qquad \qquad  \qquad \qquad \qquad \la{678}\te   A_{\ru}^{(2)} =\frac{\rt^2 \,(2\, \rs^4-10\, \rs^3\, \ru
+33\, \rs^2\, \ru^2-24\, \rs\, \ru^3+3\, \ru^4)}{7680\, \rs^4} , \\
& \pm2 \mp2 \to  \pm2\mp2: \ \ \qquad   \te A_{\rs}^{(2)} = 0, \quad \no 
A_{\rt }^{(2)} =\frac{\ru^4 \,(2\, \rs^2+2\, \rs\, \rt+3\, \rt^2)}{7680\, \rs^4} , \quad
A_{\ru}^{(2)} =\frac{\ru^4 \,(10\, \rs^2+18\, \rs\, \ru+9\, \ru^2)}{23040\, \rs^4} \ . 
\end{align}
The contributions  of the  4-derivative  2222   contact vertex  which is the $s=2$ analog of \rf{2.99} 
 %(containing 4 derivatives according to \rf{2}) 
are   found to be 
\be\la{68}
\begin{split}
&  \pm2 \pm2 \to  \pm2\pm2: \ \ \qquad A_{ }^{\rm (cont)} \te = -\frac{\rs^6-\rs^5\, \rt+26\, \rs^4\, \rt^2+63\, \rs^3 \,\rt^3
+54\, \rs^2\, \rt^4+27\, \rs \,\rt^5+9 \,\rt^6}{1920\, \rs^4}\ , \\
& \pm2 \mp2 \to  \pm2\mp2: \ \ \qquad  A_{}^{\rm (cont)} \te = -\frac{\ru^{4}\, (\rs^2+3 \,\rs \,\rt+9\, \rt^2)}{1920\, \rs^4} \ .
\end{split}
\ee
Remarkably, the sum of   these three contributions   vanishes   for each of the  helicity   
choices: 
\be\la{69}  %22 \to 22: \ \ \  \ \ \ \ \ 
[A^{(0)}_{\rs} + A^{(0)}_{\rt} + A^{(0)}_{\ru}]  + [A^{(2)}_{\rs} + A^{(2)}_{\rt}] + A^{(2)}_{\ru} 
+A^{\rm (cont)} = 0 \ . 
\ee
Note that 
this result is equivalent to the vanishing of the 4-graviton scattering 
amplitude in the non-linear $C^2$ Weyl gravity theory. 
Indeed, the  linear scalar - CHS coupling action \rf{3} is equivalent  to the  covariant  conformal scalar  action \rf{4} 
%(with non-linear couplings to graviton)
 by a local field redefinition \rf{6}. As the latter action directly leads to the Weyl tensor  squared 
 action as the "induced" one \rf{5}, 
% the cubic 222  couplings in the  CHS and 
and  as the field  redefinitions of $h_0$ and $h_2$    in the spin $\leq 2$ part of the 
 CHS action   induced from \rf{3},\rf{2.1} should not change the  graviton  S-matrix, the latter 
 should be  same as   in the Weyl theory.  In more detail,  
 adding the exchange of the  non-propagating $h_0$ field    produces (as it follows from \rf{2.111})
   an extra  4-derivative 2222  contact vertex 
 contribution.  %Integrating out $h_0$ means  that its redefinition in \rf{6} is not 
 The  remaining local  redefinition of $h_{\m\n}$  in \rf{6}  may alter  the 222  vertices by terms   proportional to the  linearised   equations of motion  ($\Box^2 h_{\m\n\ } ^{\rm TT}=0$)  and also change the  quartic 2222  vertex, but   
 it cannot change the  resulting on-shell 
  4-graviton scattering amplitude.
  
  The   vanishing of the tree-level 4-graviton amplitude  in Weyl theory  can be 
  deduced also 
  from the expressions  in \cite{Dona:2015tra}  for the massless graviton scattering in 
  $L= a R + b C^2$ theory   by taking the limit $a\to 0$   in the  final expression  for the 4-graviton amplitude. 
  The  propagator  here is symbolically $\frac{1}{a p^{2}  + b p^4}$   \ci{Stelle:1977ry} 
  (reducing to the Weyl graviton propagator for $a\to 0$ or to the Einstein propagator  for $b\to 0$) 
   so as long as the asymptotic states are chosen to be massless helicity $\pm 2$  gravitons the resulting  amplitude 
   interpolates  smoothly between  the  standard Einstein 
   4-graviton one and  zero   in the Weyl theory.\foot{Let us also mention  that 
   the conformal graviton 
amplitudes in flat space  were computed   in   \cite{Dolan:2008gc}   starting with 
the  twistor string  theory of \ci{Berkovits:2004jj}. The latter  should   be related to "non-minimal"   conformal supergravity 
containing extra  dimension 0 scalar coupling to Weyl squared term, $\phi \Box^2 \phi +  (1+  k\,  \phi + ... )C^2 + ...$. 
The tree-level 4-graviton   amplitude   in such theory is given by the sum of the 4-graviton 
amplitude in Weyl theory   and the scalar exchange $\sim  k^2   C^2 \Box^{-2}   C^2$. 
The non-zero result for the 4-graviton amplitude found in  \cite{Dolan:2008gc}  appears to be given just by this  
scalar exchange, {\em i.e.}  is consistent  with the vanishing of the graviton  amplitude in pure Weyl theory. 
Similar  result  was found  in  \cite{Adamo:2012xe,Adamo:2013tja} by 
taking the  flat limit of the conformal  graviton  scattering amplitude in  dS   space 
which is the same as the Einstein gravity one \ci{Maldacena:2011mk}  times the cosmological constant factor.}

   Remarkably, as we have just seen,  the vanishing of tree-level 4-graviton amplitude in Weyl theory
     generalises to the full CHS theory: 
   the  results \rf{66},\rf{677}    and \rf{69}   combined  together  imply 
 that  like the 11$\to$11   amplitude in \rf{3.25}  (and also the 
 external conformal scalar amplitude \ci{Joung:2015eny}) the  total 22$\to$22   conformal graviton scattering amplitude
   in the CHS theory  vanishes  after all intermediate exchange 
    contributions are added together.

%The {\bf conformal} graviton polarisations projecting onto the massless spin 2 component 
%are given by 
%$\varepsilon_{\mu\nu}=\varepsilon_{\mu}\varepsilon_{\nu}$. About Weyl graviton scattering in Weyl gravity
%we make the following remarks

\subsection{$11\to 22$  scattering} 

One may also  consider  some  "mixed"  4-particle amplitudes involving both vectors and  conformal gravitons.
The  amplitudes with odd number of vectors vanish  identically  so one is to consider only 11$\to $ 22   case. 
Here the  two   a priori non-trivial helicity choices are $\pm1\mp1\to  \pm2 \mp2$ and $\pm1\pm1\to  \pm2 \mp2$.

%%%%%%%%%%%%%%%%%%%%%%%%%%%%%%%%%%%
Let us first  briefly mention also  the expressions   for  the  "mixed"   amplitude 
 where  two external conformal  scalars $\vp$ in \rf{2.1} 
scatter into two conformal gravitons. 
As in the $\varphi\varphi^{*}\to 11$   case \rf{345}   the  non-vanishing  helicity-preserving amplitude 
 $\vp\vp^* \to \pm 2\mp 2$  receives  contributions from even spin $s\ge 4$ exchanges  that should  have the general structure 
 consistent again  with $J=s$ term in \rf{41},\rf{44}:
\be
\la{6.14}
\begin{split}
A^{(s)}_{\rs\ 0,0; \pm 2 ,\mp 2} %(\rs, \cos\theta) = 
=\te  (s+\tfrac{1}{2}) F^{(s)}_{0,0; \pm 2, \mp 2}\,\rs \,\big(\frac{\ru\,\rt}{\rs^{2}}\big)^{2}
\,P_{s-4}^{(4,4)}  \big(-1-2\,\frac{\rt}{\rs}\big) \ .  
\end{split}
\ee
The explicit computation for $c_s =(s+\tfrac{1}{2}) F^{(s)}_{0,0; \pm 2, \mp 2}$ gives again the result consistent 
with the ansatz \rf{5.1} (here $\l=0, \ \mu=4, \ M=4, N=0$) 
\be\la{613}\te 
c_{s} =(s+\tfrac{1}{2}) F^{(s)}_{0,0; \pm 2, \mp 2} = -\frac{3}{4}\,\frac{2s+1}{(s-3)(s-2)(s-1)s} \ .
\ee
To get the full amplitude one is to  add  also the contributions of the $s=0, 2$  exchanges. 

Turning  to the   $\pm1\mp1\to  \pm2 \mp2$   amplitude,    we find that  the non-vanishing helicity-preserving even $s\geq 4$ exchange 
amplitude  in the  s-channel has again  the form as   predicted by \rf{41},\rf{44} 
%(here $s\geq 4$ and even) 
\begin{align} \la{620}
&A^{(s)}_{\rs \  \ \pm1,\mp1; \pm2,\mp2} =  c_s\, \rs\,  \te 
\frac{\rt\,\ru^{3}}{\rs^{4}}\,P_{s-4}^{(6,2)}(-1-2\,\frac{\rt}{\rs}) \ , \\
& \ \ \ c_s =  (s+\tfrac{1}{2}) F^{(s)}_{\pm1,\pm 1; \pm 2, \mp 2} = \te  \frac{3}{2}\,\frac{2s+1}{(s-3)(s-2)(s-1)s(s+1)(s+2)}\ , \la{621}
\end{align}
where $\l=2, \mu=4, M=4, N=2$  so the expression for $c_s$ is again consistent with \rf{5.1}. 
In the t-channel one finds (after an appropriate relabelling of helicities and kinematic variables)
that  for  odd $s\geq 3$ \foot{Here  we  took into account that 
for odd spin the momentum space propagator has an extra factor $(-1)^s=-1$, cf. \rf{2.6}.}
\begin{align} \la{6211}
&A^{(s)}_{\rs \  \ \pm1,\mp1; \pm2,\mp2} =  c'_s\, \rs\,  \te 
\frac{\ru^{3}}{\rs \rt^{2}}\,P_{s-4}^{(6,0)}(-1-2\,\frac{\rs}{\rt}) \ , \\
& \ \ \ \ c'_s   = \te  \,\frac{2s+1}{(s-2)(s-1)s(s+1)(s+2)(s+3)}\ .  \la{62111}
\end{align}
The u-channel contribution  is zero. 
The  total  s- plus t-channel  contribution to the 
 amplitude from these higher spin exchanges   is   then  % ($x= \frac{\rt}{\rs}$)
\ba
&\qquad \qquad  A_{s>2}= {\ru^3\ov \rs^2} \bar A({\rt\ov \rs}) \ , \qquad \qquad 
 \bar A(x)= x\,S(x)+x^{-2}\,T(x^{-1}) \ ,  \la{622} \ \\
&S(x) \equiv \sum_{s=4,6,8,...}^{\infty} c_s \,
\,P_{s-4}^{(6,2)}(-1-2\,x) \ ,\   \ \ \  \ \ \ \ \ \ \ \  \ \ \   T(x) \equiv \sum_{s=3,5,7,...}^{\infty} c'_s \,
\,P_{s-4}^{(6,0)}(-1-2\,x) \ . \la{623}
\end{align}
The explicit evaluation of $S(x)$ and $T(x)$  for $-1<x<0$  gives  
\be\la{624}
\begin{split}\te 
S(x) = -\frac{x^3+5 x^2+13 x-3}{96 (x+1)^5}-\frac{x \log (-x)}{8 (x+1)^6}, \qquad \qquad 
T(x) = -\frac{(x-1) (x^2+8 x+1)}{96 (x+1)^5}-\frac{x^2 \log (-x)}{8
   (x+1)^6}\ , 
\end{split}
\ee
so that $ \bar A = -\frac{1}{96\,(x+1)}$. The  resulting contribution  of all higher $s>2$ spin exchanges  to the total 
 amplitude  is thus 
\be\la{625} \te 
A_{s>2}= -\frac{1}{96}\,\frac{\ru^{3}}{\rs^{2}}\frac{1}{\frac{\rt}{\rs} +1 } = \frac{1}{96}
\frac{\ru^{2}}{\rs} \ . 
\ee
We are  still to   add possible  contributions  of  low-spin $s=0,1,2$  exchanges and contact 1122  vertex. 
The $h_0$ exchange is trivial  as the $110$ vertex \rf{2.7}  vanishes  for the   $\pm1,\mp1$ helicity  choice. 
The $h_2$   exchange is also found to vanish. The $h_1$ exchange gives the following non-zero 
contributions in the  t- and u-channels 
\be\te \pm1\mp1 \to \pm2\mp2: \ \ \ \ \qquad 
A^{(1)}_{\rt } = \frac{\ru^{3}\,(2\rs+\rt)}{192\,\rs^{3}}, \qquad \qquad 
A^{(1)}_{\ru } = \frac{\rt\,\ru^{2}\,(\rs-\rt)}{192\,\rs^{3}}
\ . \la{628}
\ee
The 1122  contact term  can be found   by computing the UV singular part of the  scalar loop diagram 
with two spin 1 and two spin 2 current insertions  
\be\la{fig5}
\vcenter{\hbox{\includegraphics[scale=0.3]{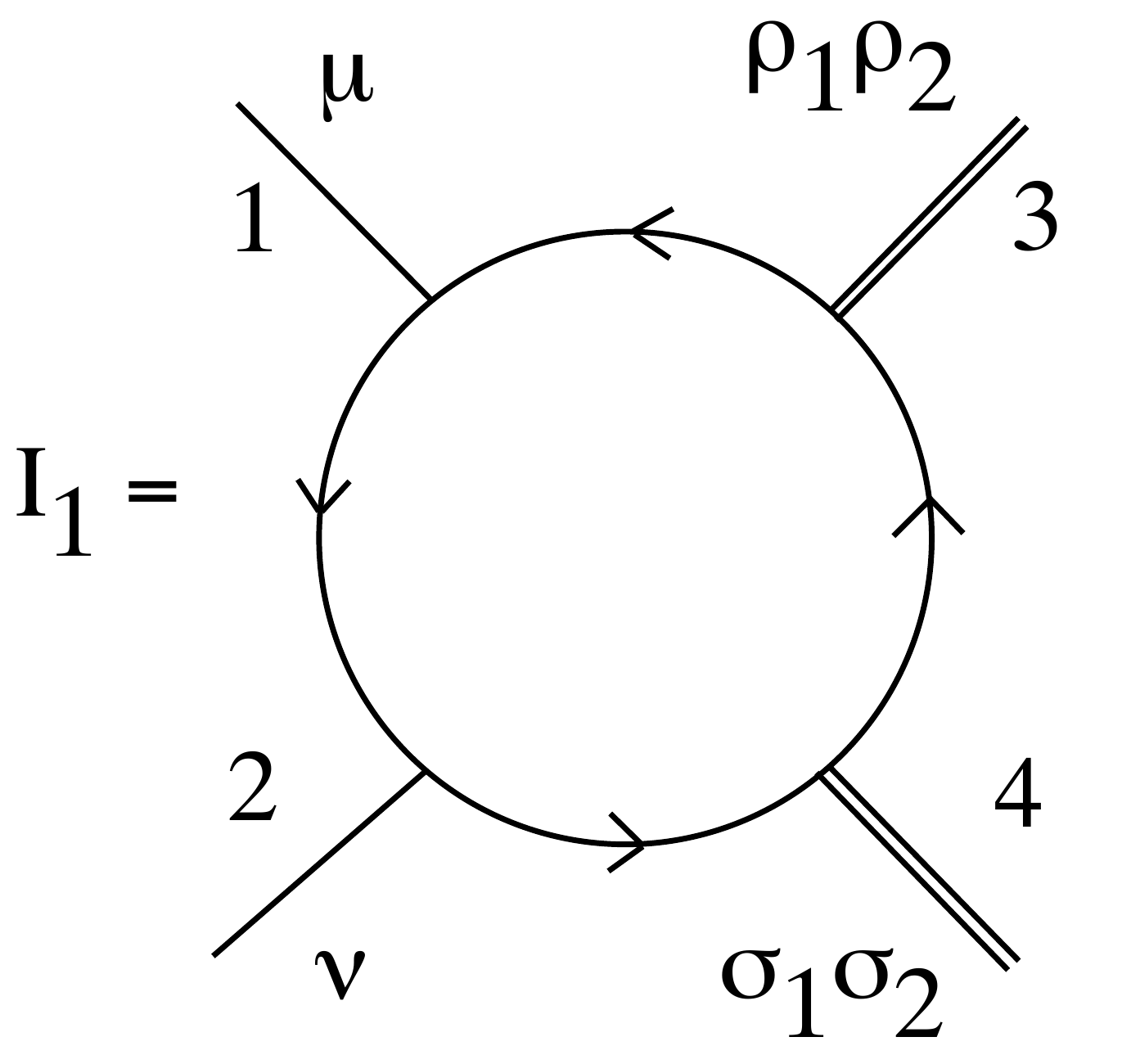}
\hskip 1cm
\includegraphics[scale=0.3]{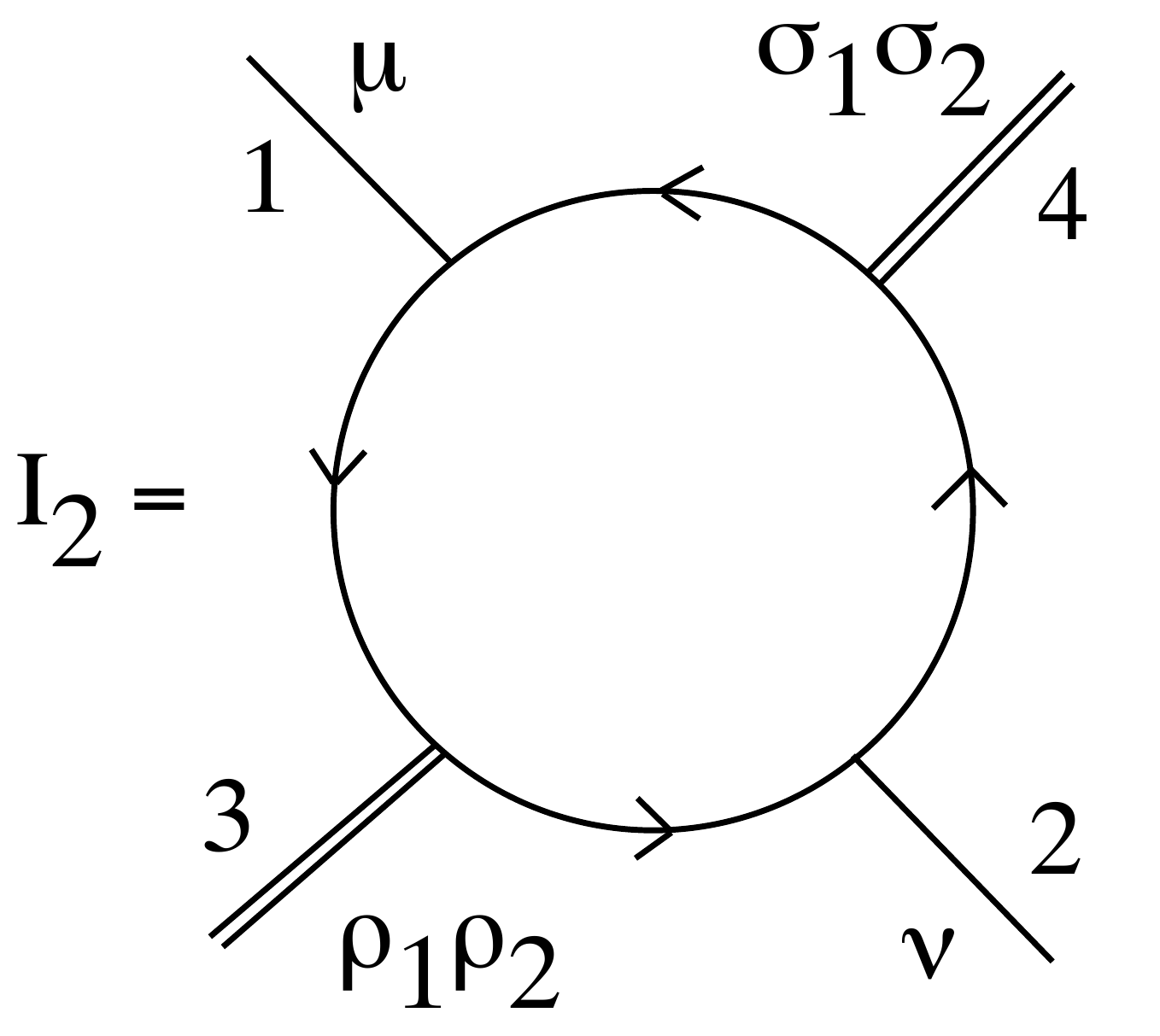}}}
\ee
 and then  replacing  the legs  with physical polarisations.\foot{Here the diagrams with the opposite loop orientation
give  equal contributions.} 
%Let us calculate directly the {\bf amplitude} by evaluating the diagrams
%and contracting with polarisations. This is same as computing the diagrams with free indices, extract the 
%Feynman rule, and using it to compute the contact CHS diagram. The only difference is an additional
%minus sign from $\exp(-S_{\rm cubic})$. 
We get 
%\be \begin{split}
%I_{1} &= \frac{1}{16\,\pi^{2}\,\varepsilon}
%\epsilon_{1\mu}\epsilon_{2\nu} \epsilon_{3\rho_{1}\rho_{2}}^{*}\epsilon_{4\sigma_{1}\sigma_{2}}^{*}
%\int \frac{d^4k}{2\pi}\big)^{d}\frac{k_{\mu}(k+p_{1})_{\nu}\cdots}{
%k^{2}\,(k+p_{1})^{2})(k+p_{1}+p_{2})^{2}(k+p_{1}+p_{2}+p_{4})^{2}}\\&+\text{finite part},
%\end{split} \ee
%and similar for $I_{2}$. Then, 
$
I_{1} = \frac{1}{20}\,\frac{\rt\,\ru^{3}}{\rs^{3}},  \ \  I_{2} =  -\frac{1}{60}\,\frac{\rt\,\ru^{3}}{\rs^{3}}$.
Including combinatorics  factors   the total   contribution is $ (-1) {1\ov (2!)^4} 2 (2I_1 +I_2)$, {\em i.e.} 
\be
\la{627}\te  \pm1 \mp1 \to \pm2 \mp2: \ \ \ \ \qquad \qquad 
A^{\rm (cont)}_{ } = -\frac{1}{96}\frac{\rt\ru^{3}}{\rs^{3}} \ .
\ee
The total   amplitude given by the sum of the contributions 
of the  higher-spin exchanges \rf{625}, spin 1 exchange \rf{628}  and  the 4-point contact  vertex \rf{627}
is found to vanish 
\be\la{629}
 \pm1\mp1 \to \pm2\mp2: \ \ \ \ \qquad A_{s>2} + \big[
  A^{(1)}_{\rt } + A^{(1)}_{\ru }\big]  + A^{(\rm cont)}=0 \ . 
%\frac{1}{96}\,\frac{\ru^{2}}{\rs}
%+\frac{\ru^{3}\,(2\rs+\rt)}{192\,\rs^{3}}+\frac{\rt\,\ru^{2}\,(\rs-\rt)}{192\,\rs^{3}}
%-\frac{1}{96}\frac{\rt\ru^{3}}{\rs^{3}} = 0 \ .
\ee 
Let us now consider the second  non-trivial   helicity amplitude $\pm1\pm1\to  \pm2 \pm2$. 
Here  we find that the s-channel   amplitude vanishes while 
  the higher odd $s\geq 3$ spin exchange contributions  in t- and u-channels  have the form 
 consistent  with the general expectations \rf{41},\rf{44},\rf{5.1} 
\ba \te 
\pm1\pm1\to \pm2\pm2\ : \qquad  \ \ \ \ \te 
&A_{\rt}^{(s)}=\te  c_s \frac{\rs^{3}}{\rt^{2}}
\,P_{s-3}^{(0,6)}(-1-2\frac{\rs}{\rt}), \ \qquad 
A_{\ru}^{(s)} = c_s \frac{\rs^{3}}{\ru^{2}}
\,P_{s-3}^{(0,6)}(-1-2\frac{\rs}{\ru})\ ,  \no \\
& \qquad  \te c_s = -\frac{2s+1}{(s-2)(s-1)s(s+1)(s+2)(s+3)} \ , \ \ \ \ \ s=3,5,7,...\  \ .   \la{640}
\end{align}
The sum over  all  odd spin $s\ge 3$  exchanges 
can be done  by  observing  that $
P_{s-3}^{(0,6)}(-1-2x) = P_{s-3}^{(6,0)}(1+2x)$  and that $c_s$ in \rf{640} is minus $c_s'$ in \rf{62111}.
One can then use the expression for $T(x)$ in \rf{623},\rf{624} to find that  the sum 
of  t- and u-channel  amplitudes in \rf{640}  vanishes  as a consequence of  ($x = {\rs\ov \rt}$)
\be\te \la{642}
x^{2}\,T(-1-x)+(\frac{x}{1+x})^{2}\,T(-\frac{1}{1+x}) = 0\ .
\ee
For the  non-vanishing low-spin exchange   and the 1122 contact term  contributions  here we get 
\ba 
\pm1\pm1\to \pm2\pm2\ : \qquad &  \te 
A_{\rs}^{(0)}= -\frac{\rs}{128}, \qquad
A_{\rt}^{(1)}= \frac{\ru^{2}\,(\rs^{2}-6\,\rs\,\rt+2\rt^{2})}{128\,\rs^{3}}, \qquad
A_{\ru}^{(1)}= \frac{\rt^{2}\,(\rs^{2}-6\,\rs\,\ru+2\ru^{2})}{128\,\rs^{3}}, \no\\
& \la{644}\te 
A^{\text{(cont)}} = -\frac{\rt\,\ru\,(\rt^{2}+3\,\rt\,\ru+\ru^{2})}{32\,\rs^{3}}.
\end{align} 
They  separately  sum up to zero
\be\la{645}
A_{\rs}^{(0)} + [A_{\rt}^{(1)}  + A_{\rt}^{(1)}]  + A^{\text{(cont)}}   =0 \ , 
\ee
so the total $\pm1\pm1\to \pm2\pm2$  amplitude is again zero. 

%New
We conclude that like the 11$\to$11 and 22$\to$22  amplitudes, the 11$\to$22  amplitudes  also vanish. 
 It is thus  natural to conjecture that all higher spin amplitudes in the CHS theory should also vanish. 
 In Appendix \ref{B}  we provide a check of this conjecture by  demonstrating that the exchange amplitude 
 for the scattering of four  spin j CHS particles  constructed  using the general relations  in \rf{41},\rf{44},\rf{5.1} 
vanishes at the special kinematical point $\ru=0$  (\ie for backward scattering).

%%%%%%%%%%%%%%%%%%%%%%%%%%%%%%%%%%%%%%%%%%%%%

\section{Concluding remarks}
%%%%%%%%%%%%%%%%%%%%%%%

In this paper we  provided  evidence that tree-level 4-particle scattering amplitudes for (the standard massless 
modes of) conformal  higher spin fields  vanish after  summing over all intermediate  CHS  exchanges. 
The amplitudes vanish  due to cancellation between 
 the summed up contributions of different scattering  channels. 
This is an indication that  this  cancellation  may be a consequence of  the underlying   higher spin symmetry
which is an infinite dimensional extension of the usual conformal symmetry. 

Indeed, the CHS theory inherits the global  higher spin symmetry of the  free  scalar  theory  (which is also a 
 symmetry of the dual massless higher spin theory in AdS$_5$). 
 This symmetry acts on  the scalar $\vp$ and the source  fields $h_s$ in the coupled action \rf{2.1} 
 and thus  becomes  the symmetry of the local UV part  of the induced  action 
 \ci{Segal:2002gd,Bekaert:2010ky}.\foot{It also acts on the 
 correlation functions of currents $J_s$ at separated points  and,   vice versa, requiring  it to be a symmetry of  these correlation functions implies that they should correspond to a free  CFT \ci{Maldacena:2011jn,Boulanger:2013zza,Stanev:2013qra,Alba:2015upa}.}
 
 As was shown  in \ci{Joung:2015eny},  the  vanishing of the   conformal scalar 4-point amplitude  in the coupled   
 scalar   -- CHS theory   can  be understood as a consequence of a particular subset of transformations of the higher spin algebra -- the hypertranslations  $\delta \vp = \eps^{\m(k)}\del_{\mu(k)} \vp$   and  the 
 rescalings. 
 Similar reasoning should apply also in the case of the scattering amplitudes with CHS fields on external lines  considered in the present paper. The transformation of the CHS   fields  $h_s$ 
 under the  differential   %and algebraic ($\a$) 
 part of the  gauge symmetry     is symbolically 
 $\delta h= \del \eps  + \eps \del h $, i.e. like the usual  diffeomorphisms it   contains an  inhomogeneous
 and homogeneous   parts  with the latter mixing different spins. The global part of  the algebra corresponds to $\eps$ being  chosen as conformal  Killing tensors. For example, for constant $\eps$ 
 \ba
& \delta  h_0 = \sum_k \e^{\m(k)}\partial_{\m(k)} h_0 \ , \ \ \ \ \ \qquad 
\delta  h^{\r} =\sum_k \big[  \e^{\r\m(k)}\partial_{\m(k)}  h_0 +  \e^{\m(k)} \partial_{\m(k)} h^{\r} \big] \ , \la{6.1} \\
&\delta  h^{\r\s}=  \sum_k \big[ \e^{\r\s\m(k)}\partial_{\m(k)}  h_0 + 2 \e^{\m(k) (\r}\partial_{\m(k)} h^{\s)}  
+ \e^{\m(k)}\partial_{\m(k)} h^{\r\s}\big] \ , ... \la{6.22} 
\end{align}
 These transformations   relate Green's functions with  different  types of legs $h_0, h_1, h_2, ...$. 
 In the case of the S-matrix  where  the non-propagating   field $h_0$ does not appear  on external lines 
 the transformation of $h_1$  under hypertranslations will be the same as of the conformal  scalar  in 
\ci{Joung:2015eny}  so that  choosing $\eps^{\m_1...\m_k}= y^{\m_1 }... y^{\m_{k}}$   where  $y^\m$ is an arbitrary constant vector   we may then repeat the argument of \ci{Joung:2015eny}  for the vanishing of the corresponding scattering amplitude. 
Similar arguments  should also  apply to amplitudes involving conformal gravitons. 
 
 As the same higher spin algebra   controls also the massless higher spin  theory in AdS$_5$ \ci{Vasiliev:2003ev}
  (of which the CHS theory is  an effective  4d   "shadow" or  corresponds to the alternative choice of the  boundary conditions 
  for higher spin fields \ci{Giombi:2013yva}) 
     it would be interesting  to know if there is  an AdS related   argument   for 
  the vanishing of the CHS  S-matrix.  %Another interesting direction is to
One may also   start with  the CHS theory defined  on AdS$_4$ or dS$_4$  and try to  generalise the arguments of 
   \ci{Maldacena:2011mk,Adamo:2013tja}  to argue that the  corresponding  S-matrix  for massless higher spin modes of the CHS fields  should  be the same as  in the  corresponding 
   massless higher spin theory. % times a factor of cosmological constant.  
  Taking   the flat limit (i.e. the  cosmological constant to zero)
   may then  lead to the conclusion  that  the S-matrix  of the resulting hypothetical massless higher spin theory in flat space should also be trivial.

\iffa
\begin{itemize}
\item About the auxiliary field $h_{0}$, things may be better if we consider free fermions instead of free bosons. With only fermions the currents are still bosonic. Their general form 
is $j_s \sim \overline\psi\partial^{s-1}\gamma\psi$ with dimension $\frac{3}{2}+s-1+\frac{3}{2} = s+2$ as in bosonic case. However, derivatives are different.
At spin 0, the two point function $\langle j_{0}j_{0}\rangle\sim p^{2}$ so we get a dynamical 
$\phi \Box \phi$. This should be the usual remark about values of $\Delta$ for the singlet in 
bosonic or fermionic case. For spin $s>0$, $\langle j_{s}j_{s}\rangle\sim p^{2s}$ as in bosonic case.
By the way, it seems quite unexplored the susy case where one starts with a free chiral field with both 
$\phi$ and $\psi$. In this case we have also half integer currents.
\end{itemize}

\iffa 
The definition of scattering amplitudes with external CHS fields of spin $s>1$ 
requires a dedicated discussion. 
\begin{itemize}
\item a general problem is the violation of unitarity. Remarkably, this is a problem that has no 
relevant -- {\em i.e.} stumbling blocks -- consequences in the perturbative treatment we are considering. 
\item a technical problem is how to adapt the LSZ formula in presence of propagators
with strong IR singularities. We bypass this discussion and define on-shell amplitudes as coefficients  in on-shell value of effective 
action written as functional of in-fields that is solutions of free equations of motion. 
{\em Useful references to be discussed/added: }
general scattering \cite{Jacob:1959at,Carmi:2011dt,Dolan:2011dv}
4-graviton scattering in Weyl gravity \cite{Dona:2015tra}
graviton helicity tensors ? \cite{Gross:1968in,Grisaru:1975bx}
no-go theorems \cite{Bekaert:2010hw,Taronna:2011kt}
CHS and twistors \cite{Adamo:2012xe,Adamo:2013tja,Haehnel:2016mlb}
	\cite{eden1967high}
\fi

%%%%%%%%%%%%%%%%%%%%%%%%%%%%%%%%%%%%%%%%%%
\acknowledgments
We  would like to thank  T. McLoughlin, R. Metsaev, D. Ponomarev   and R. Roiban  for   useful discussions.  
The work
of SN and AAT was supported by the STFC  Consolidated  grant ST/L00044X/1.
The work of AAT was  also  supported   by   the 
ERC Advanced grant No.290456   and 
by  the  Russian Science Foundation grant 14-42-00047  associated with Lebedev Institute.

\
%%%%%%%%%%%%%%%%%%%%%%%%%%%%%

\newpage
%%%%%%%%%%%%%%%%%%%%%%%%%%%
\appendix

%\section{Field redefinitions}  \la{A} 
 %%%%%%%%%%%%%%%%%%%%%%%%%%%%%%%%%%%%
  
 %%%%%%%%%%%%%%%%%%%%%%%%%%%%%%%%%%%%%%%%%%%%%%%
\section{Vertices in CHS  action from scalar loop integrals}  \la{A}

Here we shall   provide some details of  computation  of UV singular parts  of the  complex scalar loop diagrams 
with few higher spin current insertions \rf{2.1} 
 leading to  the expressions  for CHS vertices given  in section 2.  For the computation of Feynman integrals 
we shall use  the standard  relations 
\begin{align}
\la{A.1}
\int \frac{d^dk}{(2\pi)^d}\frac{(k^{2})^{a}}{(k^{2}+M^{2})^{b}} = 
\frac{\Gamma(b-a-d/2)\Gamma(a+d/2)}{(4\pi)^{d/2}\Gamma(b)\Gamma(d/2)}\,(M^2)^{d/2+a-b} \ , 
\\
\la{A.2}
\frac{1}{A_{1}\dots A_{n}} = (n-1)!\,\int_{[0,1]^{n}}d^{n}x\,\frac{\delta(x_{1}+\dots+x_{n}-1)}
{(x_{1}A_{1}+\dots +x_{n}A_{n})^{n}} \ . 
\end{align}
The expression for the scalar loop diagram in \rf{f1} with two  current operator or vertex \rf{2.5} insertions 
has the following general structure 
\be\la{a3}
\begin{split}
&\int\frac{d^dk}{(2\pi)^d}\frac{N(k,p)}{k^{2}\,(k+p)^{2}}=\int_{0}^{1}dx \int\frac{d^dk}{(2\pi)^d}\frac{N(k,p)}{[(k+x\,p)^{2}+x(1-x)\,p^{2}]^{2}}\\
&\qquad =\int_{0}^{1}dx \int\frac{d^dk}{(2\pi)^d}\frac{N(k-x\,p,p)}
{(k^{2}+M^{2})^{2}},\qquad  \qquad M^{2}=x(1-x)\,p^{2}.
\end{split}
\ee
When contracted with two TT fields $h_s$   the numerator  in \rf{a3}  takes  the following form 
\be \la{a4} \te 
N_{\mu(s)\nu(s)}(k-px,p)\to \frac{1}{(s!)^{2}} k_{\mu_{1}}k_{\nu_{1}}\dots
k_{\mu_{s}}k_{\nu_{s}}\to \frac{1}{(s!)^{2}} \frac{1}{2^{s}(s+1)}(k^{2})^{s}\,
\eta_{\mu_{1}\nu_{1}}\cdots\eta_{\mu_{s}\nu_{s}}\ . 
\ee
Then integrating over $k$    and extracting the   coefficient of the $1\ov \epsilon$ pole term  
we find 
\be \la{a5} 
\begin{split}
\SS_2[h_s]={\te  \frac{(-1)^{s}}{2^{s}\,\Gamma(2s+2)}\,}
\int\frac{d^{4}p}{(2\pi)^{4}}\,
h_{\mu(s)}(p)(p^{2})^{s}\,h^{\mu(s)}(-p)\ , 
\end{split}
\ee
which becomes  \rf{2.6} when written in coordinate representation.

The cubic vertex  for three  CHS fields is   determined by the diagram in \rf{f2}.
For the 1-1-$s$ vertex multiplied   by $h_\m(p_1)h_\n(p_2) h_{\r(s)}(-p_1-p_2)$   in the CHS action  we get\foot{Here only the UV pole part 
is  to be kept: for simplicity,  here and below we shall use   use the same   notation $V$  for the 
full integral  and  the coefficient of its pole part, {\em i.e.} 
$V \to  { 1\ov (4\pi)^2 \epsilon} V +$finite.} 
%The above diagram has the expression \footnote{The $1/2!$ is due to the two 1-0-0 vertices.}
\be\la{a6}
V_{\mu,\nu,\rho(s)}(p_1,p_2) ={\frac{1}{2!}\frac{1}{s!}}\int\frac{d^dk}{(2\pi)^d}\frac{k_{\mu}(k+p_{1})_{\nu}(k+p_{1}+p_{2})_{\rho(s)}}
{k^{2}(k+p_{1})^{2}(k+p_{1}+p_{2})^{2}}\ . 
\ee
There is also another diagram with the scalar $U(1)$ charge flowing in the opposite direction
giving $\widetilde  V_{\mu,\nu,rho(s)}(p_1,p_2)= V_{\nu,\mu,\rho(s)}(p_2,p_1)$; their  sum 
\be \V_{\mu,\nu,\rho(s)}(p_1,p_2)=  V_{\mu,\nu,\rho(s)}(p_1,p_2) + V_{\nu,\mu,\rho(s)}(p_2,p_1) \la{a7} \ee
ensures the symmetry under  $h_{\mu}(p_{1})\leftrightarrow
h_{\nu}(p_{2})$. To compute the pole part of this  integral we  use
  Feynman parametrisation (and shifts of $k$)  and assume that  the external legs are  contracted with TT fields
  ({\em i.e.} terms with $p_{1\,\mu}, p_{2\nu}, (p_{1}+p_{2})_{\rho(s)}$ can be dropped).  Then 
\be\la{a8}
\begin{split}
V_{\mu,\nu,\rho(s)}&= \frac{1}{2\,s!}\int\frac{d^dk}{(2\pi)^d}\frac{k_{\mu}k_{\rho(s)}
(k+p_{1})_{\nu}}
{k^{2}(k+p_{1})^{2}(k+p_{1}+p_{2})^{2}} \\
%&\stackrel{\rm Feyn}{=} \frac{1}{s!}\int_{0}^{1}dx\int_{0}^{1-x}dy \int\frac{d^dk}{(2\pi)^d}\frac{k_{\mu}k_{\rho(s)}(k+p_{1})_{\nu}}
%{[(1-x-y) k^{2}+x (k+p_{1})^{2}+y (k+p_{1}+p_{2})^{2}]^{3}} \\
&= \frac{1}{s!}\int_{0}^{1}dx\int_{0}^{1-x}dy \int\frac{d^dk}{(2\pi)^d}
\frac{k_{\mu}k_{\rho(s)}(k+p_{1})_{\nu}}
{[(k+xp_{1}+y(p_{1}+p_{2}))^{2}+M^{2}]^{3}}\ 
\\  
&\to 
\frac{1}{s!}\int_{0}^{1}dx\int_{0}^{1-x}dy \int\frac{d^dk}{(2\pi)^d}
\frac{(k-y p_{2})_{\mu}(k-x p_{1})_{\rho(s)}(k+(1-x-y)p_{1})_{\nu}}
{(k^{2}+M^{2})^{3}} \ , 
\end{split}\ee
\be 
M^{2} = x(1-x) p_{1}^{2}+y(1-y)(p_{1}+p_{2})^{2}-2xy p_{1}\cdot(p_{1}+p_{2}) \ . \la{a88} \ee
Separating terms of different order in $k$ in the numerator,  integrating over $k$ and then over $x,y$ 
 we  find for the pole part 
\ba\no 
V_{\mu,\nu,\rho(s)} (p_1,p_2) &\te = \frac{1}{2(s+2)!}\Big\{
\eta_{\mu\nu}(p_{1})_{\rho(s)}
-\eta_{\mu\rho_{1}}p_{1\nu}p_{1\rho_{2}}\dots p_{1\rho_{s}}
+\eta_{\nu\rho_{1}}p_{2\mu}p_{1\rho_{2}}\dots p_{1\rho_{s}}\\
&\te \qquad \qquad\qquad \qquad  -\eta_{\mu\rho_{1}}\eta_{\nu\rho_{2}}
\,p_{1\rho_{3}}\dots p_{1\rho_{s}}\big[p_{1}\cdot p_{2} + \frac{s}{2}\,(p_{1}^{2}+p_{2}^{2})\big]
\Big\}\ . \la{a9}
\end{align}
The full cubic 1-1-$s$ vertex is then given by \rf{3.3} or \rf{2.66} in coordinate representation. 

The quartic 1111  vertex \rf{2.99}  is found   from the pole part of the diagram in \rf{f3} %given by 
\be\la{a10}
\begin{split}
&{\te \frac{1}{4!}\times 6\times 3!}\int_{0}^{1}dx\int_{0}^{1-x}dy\int_{0}^{1-x-y}dz \int\frac{d^dk}{(2\pi)^d}\frac{k_{\mu}k_{\nu}k_{\rho}k_{\sigma}}
{(k^{2}+M^{2})^{4}}\\
%& = \frac{1}{4!}\times 6\times 3!\times\frac{1}{24}\,(\eta_{\mu\nu}\eta_{\rho\sigma}
%+\eta_{\mu\rho}\eta_{\nu\sigma}+\eta_{\mu\sigma}\eta_{\nu\rho})\, \\
%&\qquad \int_{0}^{1}dx\int_{0}^{1-x}dy\int_{0}^{1-x-y}dz \int\frac{d^dk}{(2\pi)^d}\frac{(k^{2})^{2}}
%{(k^{2}+M^{2})^{4}} \\
&\to \ \te   \frac{1}{16\pi^{2}\epsilon}\ \frac{1}{48}\,(\eta_{\mu\nu}\eta_{\rho\sigma}
+\eta_{\mu\rho}\eta_{\nu\sigma}+\eta_{\mu\sigma}\eta_{\nu\rho}).
\end{split}
\ee 
To find the  2-2-$s$    vertex   which is multiplied    by the TT fields    $h_{\m_1\m_2}(p_1) h_{\n_1 \n_2} (p_2)   h_{\r(s)}$ 
in the  CHS action we are to find  again the  singular part of the diagram \rf{f2} with the  vertices \rf{2.5}
leading to the integral (where we are allowed to drop TT-trivial terms proportional to $p_{1\,\mu_1}, p_{1\,\mu_2}$, etc.)
\be\la{a11}
\begin{split}
V_{\mu_{1}\mu_{2},\nu_{1}\nu_{2},\rho(s)}&=\frac{1}{2!}\frac{1}{(2!)^{2}}\frac{1}{s!}\int\frac{d^dk}{(2\pi)^d}\frac{k_{\mu_{1}}k_{\mu_{2}}(k+p_{1})_{\nu_{1}}(k+p_{1})_{\nu_{2}}(k+p_{1}+p_{2})_{\rho(s)}}
{k^{2}(k+p_{1})^{2}(k+p_{1}+p_{2})^{2}} \\
&\to \  \frac{1}{4\,s!}\int_{0}^{1}dx\int_{0}^{1-x}dy \int\frac{d^dk}{(2\pi)^d}
\frac{k_{\mu_{1}}k_{\mu_{2}}k_{\rho(s)}
(k+p_{1})_{\nu_{1}}(k+p_{1})_{\nu_{2}}}
{[(k+xp_{1}+y(p_{1}+p_{2}))^{2}+M^{2}]^{3}}\\
& \to \frac{1}{4\,s!}\int_{0}^{1}dx\int_{0}^{1-x}dy \int\frac{d^dk}{(2\pi)^d}
\frac{   N_{\mu_{1}\mu_{2},\nu_{1}\nu_{2},\rho(s)}(p_1,p_2,k; x,y)}
{(k^{2}+M^{2})^{3}} \ , 
\end{split}
\ee
where $M^2$ is the same as in \rf{a88} and 
\ba
  &N_{\mu_{1}\mu_{2},\nu_{1}\nu_{2},\rho(s)} =    (k-y p_{2})_{\mu_{1}}(k-y p_{2})_{\mu_{2}}(k-x p_{1})_{\rho(s)}
[k+(1-x-y)p_{1}]_{\nu_{1}}[k+(1-x-y)p_{1}]_{\nu_{2}}\no %\\ &\qquad 
%M^{2} = x(1-x) p_{1}^{2}+y(1-y)(p_{1}+p_{2})^{2}-2xy p_{1}\cdot(p_{1}+p_{2}).
\la{a12}
\end{align}
Non-trivial  UV  divergent pole contributions  may come from the 
 terms  in $N$  which are of order $k^{2}, k^{4}, k^{6},k^{8}$
(integrals of higher  powers of $k$  will lead to contractions between $\rho$ indices that  can be discarded due to TT condition). 
As a result, we find the vertex given in \rf{2.10}. 

Let us  also  discuss   some  vertices involving the non-propagating spin 0 field $h_0$. 
One can show that 0-0-$s$  interaction is  absent if $h_s$ is  subject to TT condition. 
1-0-$s$ vertex is given by the pole part of the integral (here $p=p_2$  and $M^2$ is as in \rf{a88})
\be\la{a13}
\begin{split}
\text{V}_{\mu,\rho(s)}&=
\frac{4}{s!}\int_{0}^{1}dx\int_{0}^{1-x}dy \int\frac{d^dk}{(2\pi)^d}
\frac{(k-y p)_{\mu}(k+x p)_{\rho(s)}}
{(k^{2}+M^{2})^{3}} %\ , \\
% M^2 &= x(1-x) p_{1}^{2}+y(1-y)(p_{1}+p_{2})^{2}-2xy p_{1}\cdot(p_{1}+p_{2})
 \ . 
\end{split}
\ee
It is non-vanishing for odd $s$ and reduces to \rf{2.11}. 
In the case of  2-0-$s$  vertex we get two diagrams \rf{f2}  with opposite   loop direction  and the sum 
of the corresponding  integrals  can be put into  the form  
\be
\la{a14}
\begin{split}
\text{V}_{\mu_{1}\mu_{2},\rho(s)}&=\frac{1}{s!}\int\frac{d^dk}{(2\pi)^d}\frac{k_{\mu_{1}}k_{\mu_{2}}k_{\rho(s)}}
{k^{2}(k+p_{1})^{2}(k+p_{1}+p_{2})^{2}} \\
&\to \  \frac{2}{s!}\int_{0}^{1}dx\int_{0}^{1-x}dy \int\frac{d^dk}{(2\pi)^d}
\frac{k_{\mu_{1}}k_{\mu_{2}}k_{\rho(s)}}
{[(k+xp_{1}+y(p_{1}+p_{2}))^{2}+M^{2}]^{3}}\\
&\to \ \frac{2}{s!}\int_{0}^{1}dx\int_{0}^{1-x}dy \int\frac{d^dk}{(2\pi)^d}
\frac{(k-y p)_{\mu_{1}}(k-y p)_{\mu_{2}}(k+x p)_{\rho(s)}}
{(k^{2}+M^{2})^{3}}\ , 
\end{split}
\ee
where $M^2$ is as in \rf{a88}. The  pole part of this integral  is given by  \rf{2.12}. 
The computation of the 1-2-$s$ vertex is similar, leading to the expression in  \rf{2.13}.

%%%%%%%%%%%%%%%%%%%%%%%%%%%%%%%%%%%%%%%
\section{Vanishing of 4-particle   amplitude at special kinematics}
\la{B}
%%%%%%%%%%%%%%%%%%%%%%%%%%%%%%%%%%%%%%%%

\subsection{$11\to 11$  scattering }

As a  check of the  vanishing of the summed over spins $11\to 11$    scattering amplitude    discussed in  section 3.2 
here we independently demonstrate this   at the  special kinematics point $\ru=0$ or $\rs=-\rt$ 
corresponding  to $\theta=\pi$ or backward scattering. The total   $++\to ++$ 
amplitude obtained  by summing \rf{3.16} over all   even spins $s=2,4,...$ for $\rs=-\rt$ may be written as 
\be\la{b1} 
 \sum_{s=2,4,...}^{\infty} A^{(s)}\Big|_{\ru\to 0} = a +  \lim_{\g \to \infty}  f(\g)  \ , \qquad 
a= {\sum_{s=2,4,...}^{\infty}}  c_{s}\ , \qquad  f(\g)  = {\sum_{s=2,4,...}^{\infty}}  c_{s}\, \g^{s}\, P_{s}(0)\,   \ , 
\ee
where the first term comes from the  t-channel and the second  from 
 the u-channel contribution (where we already set $\ru=0$ in the argument of $P_s$). 
Here we  defined  $\g\equiv {\rs\ov \ru}\to \infty$  and used   that according to \rf{3.14}  $P_s (-1)=1$.
From  \rf{3.15} we  have  
\be 
a = \sum_{s=2,4,...}^{\infty}\te 
\frac{2\,s+1}{2\,(s-1)\,s\,(s+1)\,(s+2)} = \frac{1}{8} \ . \la{b2}
\ee
From \rf{3.14}  and  \rf{3.15}  we get 
\be\la{b3}
c_{s}\,P_{s}(0) \te = \frac{\Gamma(2s+2)}{2\,\big[\Gamma(s+3)\big]^{2}}  \ , 
\ee
 and then find  that 
\be\la{b4}
\begin{split}
f(\g) &\te  % \sum_{s=2}^{\infty}  \g^{s}\,c_{s}\,P_{s}(0) 
= -\frac{1}{48\,\g^{2}}\,
\Big[6 \g^2 \  _4F_3\big(\frac{3}{4},1,1,\frac{5}{4};\frac{3}{2},2,2;16 \g^2\big)
+160
   \g^4 \  _3F_2\big(\frac{3}{2},\frac{7}{4},\frac{9}{4};\frac{5}{2},\frac{5}{2};16
   \g^2\big)\\ & \te\qquad\qquad \qquad   -3 \sqrt{2} \sqrt{\sqrt{1-16 \g^2}+1}
   +6 \big(\g-\frac{1}{\sqrt{1-4
   \g}}+\frac{1}{\sqrt{4 \g+1}}\big) \g+6\Big].
   \end{split}
\ee
%This is the generating function of the small $\g$ power series 
For small $\g$ we get the expansion 
%\be\la{b5}
%\mathop{\sum_{s=2}^{\infty}}_{s\ \rm even}  \frac{\Gamma(2s+2)}{2\,\Gamma(s+3)^{2}} \,\g^{s} 
$ f(\g) \te = \frac{5 \g^2}{48}+\frac{7 \g^4}{20}+\frac{429 \g^6}{224}+... %\frac{2431 \g^8}{180}+\frac{29393
 %  \g^{10}}{264}+\frac{185725 \g^{12}}{182}+\dots.
$  
which  is convergent for $|\g|<\frac{1}{4}$. Using the analytic continuation, we may 
extend $f(\g)$  beyond this convergence disk and evaluate it at $\g\to \infty$,  getting 
%The limit is finite and reads
\be\la{b5} 
\lim_{\g\to\infty}f(\g) \te = -\frac{1}{8} \ . 
\ee
As a result, we conclude that the $\ru\to 0$ limits of the t-channel \rf{b2}  and u-channel  \rf{b5} 
contributions  indeed cancel against each other  just 
 as  was found   for   general kinematics in \rf{3.25}.

%%%%%%%%%%%%%%%%%%%%%%%%
\subsection{ j j $ \to$\,  j j \ scattering }

To  check  our conjecture  that all 4-point CHS  amplitudes   should vanish 
we may  (i)   first make a guess for the CHS spin $\j $ 4-point  exchange amplitude  generalising the expressions 
for  the 11$\to$11   and 22$\to$22 
amplitudes explicitly computed in \rf{3.16},\rf{3.14}    and \rf{6.3},\rf{63}   being guided by the  expected structure of spin $J\geq 2\j  $ exchange amplitude  in \rf{41},\rf{44},\rf{5.1}  and  (ii) 
then check its vanishing at a special kinematical point.\foot{
%New 
We shall assume  that as in the $\j =1$ and $\j =2$ cases (see section 3.2  and \rf{69}) 
   the sum of the low-spin $J < 2\j $  exchanges and   contact $\j \j \j \j $ contribution   vanishes  separately.} 

Then the  total amplitude   is expected  to be given as in \rf{3.17},\rf{3.25}   and \rf{66},\rf{677}  by the sum of the t-channel and u-channel exchanges of even spin $s$ CHS  states %(here we ignore an overall constant factor)
\ba
\la{b6}
&\quad A= \rs^{2\j -2} \big[\sigma(x) + \sigma(-1-x) \big]  \ ,\qquad \qquad x={\rt\ov\rs} \ ,  \\
&\sigma (x) = \frac{2}{x^{2}}\, \sum_{J=2\j , 2\j +2,  ...}^{\infty}\te (J+{1\ov 2} )\,\frac{(J-2\j )!}{(J+2\j )!}\,P_{J-2\j }^{(4\j ,0)}\big(
\frac{x+2}{x}\big)\no \\ 
&\qquad = \frac{2}{x^{2}}\, \sum_{s=0,2,4, ...}^{\infty}\te (s+2\j +{1\ov 2} )\,\frac{s!}{(s+4\j )!}\,P_{s}^{(4\j ,0)}\big(
\frac{x+2}{x}\big)
\ . \la{b7}
\end{align}
Let us now show the  vanishing of the  sum $\sigma(x) + \sigma(-1-x)$ at   $\ru=0$ or $x=
{\rt\ov\rs} =-1$, {\em i.e.} $\sigma(0) = - \sigma(-1)$. This is equivalent also  to 
proving  the vanishing of this sum at $x=0$. \footnote{Notice that it is not possible to check numerically the vanishing of the amplitude 
$A$ in \rf{b6} at a generic value of the kinematical variable $x$. This is because 
the series in  \rf{b7} converges for $x\le -1$ and a test of the condition $A=0$ requires the 
analytical continuation of the series definition of $\sigma(x)$. 
%This  is provided by 
The explicit summation over the spin $s$ leads to  the result  which 
has an expected non-trivial analytical structure with branch points at $x=0, 1$, 
see, for instance, \rf{3.24} and \rf{6.6}.}

Using that $P_{s}^{(4\j ,0)}(-1) = 1$ we get\foot{Here $\pFtildeq{p}{q}{a_1...a_n}{b_1....b_m}{z}= 
{1\ov \Gamma(b_1)... \Gamma(b_m)}\ \pFq{p}{q}{a_1...a_n}{b_1....b_m}{z}$ is the regularised  hypergeometric function. We use the compact notation $\pFq{p}{q}{a_1...a_n}{b_1....b_m}{z}\equiv  \  _pF_q( a_1,...,a_n; b_1,...,b_m;z          )$ 
for  the generalised hypergeometric function.}
\be
\la{b8}
\begin{split}
\sigma(-1) &= \sum_{s=0,2,4,..}^{\infty}\te (2s+4\j +1)\,\frac{s!}{(s+4\j )!}\\ & \te = \sqrt{\pi }\, 2^{-4 \j }\, \Big[(4 \j +1) \, 
\pFtildeq{3}{2}{\frac{1}{2},1,1}{2\j +\frac{1}{2},2 \j +1}{1}
+2 \,
\pFtildeq{3}{2}{\frac{3}{2},2,2}{2\j +\frac{3}{2},2 \j +2}{1}
\Big] \\
& \te = \frac{1}{4\,(2\j -1)^{2}\,\Gamma(4\j -2)}\ .
\end{split}
\ee
To compute $\sigma(0)$  we note that  for $x\to 0$ the leading term  in the expansion of the Jacobi polynomial is %\comment\red{\j ust to tell I checked B9}
\be\la{b9}
{\te P_{s}^{(4\j ,0)}\big(
\frac{x+2}{x}\big) }\Big|_{x\to 0} = \frac{1}{x^{s}}\,\frac{4^{s+2\j }}{\sqrt\pi}\,\frac{(s+2\j )!}{s!}\,
\frac{\Gamma(s+2\j +\frac{1}{2})}{\Gamma(s+4\j +1)}+\cdots.
\ee
Plugging this into \rf{b7}    and taking the $x\to 0$ limit  we get
\be\la{b10} 
\begin{split}
\sigma(0) & \te =
\lim_{x\to 0} \Big[\te   \frac{1}{ \Gamma (4 \j )}  x^{-2} \
\pFq{5}{4}{1,\j +\frac{1}{4},\j +\frac{1}{2},\j +\frac{3}{4},\j +1}{2 \j +\frac{1}{2},2
   \j +\frac{1}{2},2 \j +1,2 \j +1}{\frac{16}{x^2}}\\
   &\qquad \qquad \te +\frac{4^{2 \j +3} \Gamma(2
   \j +  \frac{5}{2}) \Gamma (2 \j +3) }{ \sqrt \pi \, \Gamma(4\j +3) } \  x^{-4} \ \ 
   \pFq{5}{4}{2,\j +\frac{5}{4},\j +\frac{3}{2},\j +\frac{7}{4},\j +2}{2 \j +\frac{3}{2},2
   \j +\frac{3}{2},2 \j +2,2 \j +2}{\frac{16}{x^2}}
   \Big]
   \end{split}
\ee
This $x\to 0$  limit exists and can be explicitly  evaluated  for  $\j =1, 2, 3,...$. One finds  
 that  the resulting  value of \rf{b10}  is minus that of \rf{b8},\foot{It should be possible 
to prove this fact analytically given the simplicity of the result in \rf{b8}.}
  {\em i.e.}    $\sigma(0)  = - \sigma(-1)$.  
Thus the total amplitude \rf{b6}   vanishes at $\ru=0$ (or $\rt=0$).

%%%%%%%%%%%%%%%%%%%%%%%%%%%%%%%%%%%%%%%
\section{Derivation of  the general form of the 11$\to$11   spin $s$ exchange amplitude}
\la{C}
%%%%%%%%%%%%%%%%%%%%%%%%%%%%%%%%%%%%%%%%

Our starting point is the 1-1-s vertex in \rf{3.3},\rf{a9} that   may be written as 
\begin{align}
\V^{\m,\n, \r(s)}(p,q) &= \te \frac{1}{(s+2)!} q^{\r_3} \ldots q^{\r_s} \hat{V}^{\m,\n,\r_1\r_2}(p,q) \ , \no
\\
\hat{V}^{\m,\n,\r_1\r_2}(p,q) &\te = \eta^{\m\n} q^{\r_1}q^{\r_2}
 - \h^{\r_1\m} q^{\n}q^{\r_2}  +  \h^{\r_1\n} p^{\m}q^{\r_2}
 - \big[p \cdot q + \frac{s}{2}( p^2 +q^2) \big] \h^{\m\r_1}\h^{\n\r_2}  \la{c1} \ , 
\end{align}
where  symmetrisation over $\r_1,...,\r_s$ is assumed. 
Let us   contract the $\r_i$ indices  with   an  auxiliary vector $u^{\r}$, \ie define 
\begin{align}
V^{\m,\n}(p,q,u) &=\te  \frac{1}{2(s+2)!} (q \cdot u)^{s-2} \hat{V}^{\m,\n}(p,q,u) \la{c2} \ , 
\\
\hat{V}^{\m,\n}(p,q,u) &=\te  \h^{\m\n} (q\cdot u)^2 - (u^{\m} q^{\n} - u^\n p^\m )\, q\cdot u   +  
 - \big[p \cdot q +\frac{s}{2}( p^2 +q^2) \big] u^{\m} u^{\n} \ . \no 
\end{align}
The  TT projector  in the  CHS propagator \rf{3.4} acting on  monomials  of  $u$ may be  written as 
\begin{align}
& \qquad \qquad \P^{(s)}(\partial_{u_1},\partial_{u_2},k) = \frac{1}{(s!)^2} \sum_{l=0}^{[s/2]} \, a_{s,l} \, Y_1^l Y_2^l X^{s-2l} \ , \la{c3} 
\qquad \qquad 
\\ 
%\ , \qquad 
&\te a_{s,l} =(-1)^l \frac{s!\ \G(s-l+{1\ov 2} )}{ 2^{2l} l!\ (s-2l)!\ \Gamma(s+ {1\ov 2} ) } 
\ , \qquad  X \te = \partial_{u_1} \cdot \partial_{u_2} - \frac{(\partial_{u_1} \cdot k)( \partial_{u_2} \cdot k) }{k^2}
\ , \qquad 
Y_i =\partial_{u_i}^2 - \frac{(\partial_{u_i} \cdot k)( \partial_{u_i} \cdot k) }{k^2}  \ . \no 
% \qquad \quad Y_2 = \partial_{u_2}^2 - \frac{(\partial_{u_2} \cdot k )(\partial_{u_2} \cdot k) }{k^2} 
\end{align}
To compute the exchange amplitude, e.g., in the case of 
$+- \to +-$   scattering  in s-channel (cf. \rf{3.11}) 
we need to contract two vertices with CHS   propagator, i.e. compute
%The main challenge in computing the 1-1-1-1 amplitude is to understand of the TT projector acts on the two vertices, ie. understand the object:
\begin{equation}
\P^{(s)}(\partial_{u_1},\partial_{u_2},p_1+p_2) \, V^{\m_1, \m_2}(p_1,p_2,u_1)\, 
V^{\n_1, \n_2}(p_3,p_4,u_2)\Big|_{u_i=0}\ . \la{c4}
\end{equation}
Using \rf{c4}  the  $s$ dependence  is determined by 
\begin{equation}
{\P}^{(s)}_2 %(\partial_{u_1},\partial_{u_2},p_1,p_2,p_4)
 \equiv [s(s-1)]^2\, \P^{(s)}(\partial_{u_1},\partial_{u_2},p_1+p_2) \ (p_2 \cdot u_1)^{s-2}(p_4 \cdot u_2)^{s-2}\Big|_{u_i=0} \ . \la{c5}
\end{equation}
Here the subscript denotes that this is a differential operator of order $2$ in $\partial_{u_1}$ and $\partial_{u_2}$. Let us  introduce:
%v3 
\begin{align}
W_1 &\te= p_4 \cdot \partial_{u_1} + \frac{1}{2}(p_1 +p_2) \cdot \partial_{u_1} \ ,  \qquad \qquad W_2 =p_2 \cdot \partial_{u_2} - \frac{1}{2}(p_1 +p_2)\cdot \partial_{u_2} \no \ , 
\\
Z_1 &\te = %2 p_2\cdot \partial_{u_1}
 - \ha (p_1 +p_2)\cdot \partial_{u_1}\ , \qquad  \qquad Z_2 = 
  p_4 \cdot \partial_{u_2} -\ha  (p_1 +p_2) \cdot \partial_{u_2} \ , \la{c6}
\end{align}
so that we get 
%v3
%We can state the following commutations:
\begin{align}
&\big[X, (u_1 \cdot p_2)(u_2 \cdot p_4)\big] =  \tilde{\rt} + (p_2 \cdot u_1) W_2 + (p_4 \cdot u_2) W_1 
\ , \no  \\
&%\big[Y_1, (u_1 \cdot p_2)^2\big] = \tilde{\rs} + (p_2 \cdot u_1 ) Z_1  \ , 
\big[Y_1, (u_1 \cdot p_2)\big] =  Z_1  \ , 
%\qquad \Big[Y_2, (p_4 \cdot u_2)^2\big] = \tilde{\rs} + (p_4 \cdot u_2 ) Z_2 
\qquad \Big[Y_2, (p_4 \cdot u_2)\big] = Z_2 \ , 
\la{c7} \\
&
\big[Z_1, (p_2 \cdot u_1)\big]= \tilde{\rs} \ ,    \quad \Big[Z_2, (p_4 \cdot u_2)\Big]= \tilde{\rs} \ , 
\quad
\big[W_1, (p_2 \cdot u_1)\big]= \tilde{\rt}\ ,   \quad \big[W_2, (p_4 \cdot u_2)\big]= \tilde{\rt}\no 
\end{align}
where 
%v3
$\tilde{\rs} \equiv  \frac{\rs}{4} $ and $\tilde{\rt} \equiv  \frac{1}{2}\left( \rt + \frac{\rs}{2} \right) $.
Then we   may  commute $Y_1^l Y_2^l X^{s-2l}$ with $(p_2 \cdot u_1)^{s-2}(p_4 \cdot u_2)^{s-2}$  and the result will be composed of the operators $Y_1,Y_2,X,W_1,W_2,Z_1$ and $Z_2$. A  generating function
determining the   combinatoric coefficient   is % to get the correct coefficient is
\begin{align}
&\mathcal{P}^{(s)} = \sum_{j=0}^{\infty} (t_1 t_2)^{j} \ {\P}^{(s)}_{j} \no
= \sum_{l=0}^{[ \frac{s}{2}]} a_{s,l} \, \tilde{\rt}^{s-2l} \, \tilde{\rs}^{2l}
 \big[1+  \tilde{\rt}^{-1}( t_1 W_1 + t_2 W_2 + t_1 t_2 X )\big]^{s-2l}\\
& \qquad \qquad \qquad \qquad \qquad \qquad \times \big[1+ \tilde{\rs}^{-1} ( t_1 Z_1 + t_1^2 Y_1)\big]^{l}
\big[1+ \tilde{\rs}^{-1} (t_2 Z_2 + t_2^2 Y_2 ) \big]^{l} \la{c8}\ . 
\end{align}
Using this to compute \rf{c5}  and thus \rf{c4} we may get an  expression for the amplitude 
  in terms  of a hypergeometric function of the  kinematic variables. Adding the s-channel propagator $\rs^{-s}$ factor in \rf{3.4} the result  for the $+- \to  +-$   exchange amplitude  in the 
s-channel 
%for the scattering process with helicities $\l_1\l_2 \to \l_3\l_4$ (with $\l_i=\pm 1$) 
 may be written as  
\begin{equation}
A_{\rs\ +,-; +,-}^{(s)} =  \te  c_s \,  x^{-2}\,  P_{s-2}^{(4,0)}\big(\frac{x+2}{x} \big) \ , \ \ \ \  \qquad 
c_s = 2 (2s+1) { (s-2)!\ov (s+2)!} \ , \ \ \qquad    x= { \rs\ov \ru} \ ,  \la{c9} 
\end{equation}
 in agreement with \rf{3.14},\rf{3.15}.

Similar derivation may be given for the spin $s$ exchange contribution to the 22 $\to $ 22   amplitude. 
Here we  use that the 2-2-$s$ vertex \rf{2.10}  may be  written as 
\begin{align}
\V^{\m_1 \m_2,  \n_1 \n_2,  \r(s)}(p,q) = \te 
\frac{1}{8(s+4)!} q^{\r_5} \ldots q^{\r_s} \hat {V}^{\m_1 \m_2,  \n_1 \n_2,  \r(4)}(p,q) \ , \la{c10} 
\end{align}
%Once again, we write it using $u$'s:
or, when  contracted with $u^\r$, 
\begin{align}
\V^{\m_1 \m_2,  \n_1 \n_2,  \r(s)}(p,q,u) =\te 
 \frac{1}{8(s+4)!} (q\cdot u)^{s-4} \hat{V}^{\m_1 \m_2,  \n_1 \n_2 , \r(4)}(p,q,u) \ . \la{c11} 
\end{align}
Here   we need  to find  the ${\P}^{(s)}_4$  analog of ${\P}^{(s)}_2$ in \rf{c5}
and it  can  be readily  obtained from \eqref{c8}. The final result  matches 
the expressions in \rf{6.3}--\rf{65}.

\

\newpage

\bibliography{N2-Biblio}

\providecommand{\href}[2]{#2}\begingroup\raggedright\begin{thebibliography}{10}

\bibitem{Fradkin:1985am}
E.~S. Fradkin and A.~A. Tseytlin, \emph{{Conformal supergravity}},
  \href{http://dx.doi.org/10.1016/0370-1573(85)90138-3}{\emph{Phys.Rept.} {\bf
  119} (1985) 233--362}.

\bibitem{Fradkin:1989md}
E.~S. Fradkin and V.~Y. Linetsky, \emph{{Cubic interaction in conformal theory
  of integer higher spin fields in four-dimensional space-time}},
  \href{http://dx.doi.org/10.1016/0370-2693(89)90120-2}{\emph{Phys.Lett.} {\bf
  B231} (1989) 97}.

\bibitem{Tseytlin:2002gz}
A.~A. Tseytlin, \emph{{On limits of superstring in $AdS_{5}\times S^{5}$}},
  \href{http://dx.doi.org/10.1023/A:1020646014240}{\emph{Theor.Math.Phys.} {\bf
  133} (2002) 1376--1389}, [\href{http://arxiv.org/abs/hep-th/0201112}{{\tt
  hep-th/0201112}}].

\bibitem{Segal:2002gd}
A.~Y. Segal, \emph{{Conformal higher spin theory}},
  \href{http://dx.doi.org/10.1016/S0550-3213(03)00368-7}{\emph{Nucl.Phys.} {\bf
  B664} (2003) 59--130}, [\href{http://arxiv.org/abs/hep-th/0207212}{{\tt
  hep-th/0207212}}].

\bibitem{Bekaert:2010ky}
X.~Bekaert, E.~Joung and J.~Mourad, \emph{{Effective action in a higher-spin
  background}}, \href{http://dx.doi.org/10.1007/JHEP02(2011)048}{\emph{JHEP}
  {\bf 1102} (2011) 048}, [\href{http://arxiv.org/abs/1012.2103}{{\tt
  1012.2103}}].

\bibitem{Beccaria:2014jxa}
M.~Beccaria, X.~Bekaert and A.~A. Tseytlin, \emph{{Partition function of free
  conformal higher spin theory}},
  \href{http://dx.doi.org/10.1007/JHEP08(2014)113}{\emph{JHEP} {\bf 08} (2014)
  113}, [\href{http://arxiv.org/abs/1406.3542}{{\tt 1406.3542}}].

\bibitem{Haehnel:2016mlb}
P.~Haehnel and T.~McLoughlin, \emph{{Conformal Higher Spin Theory and Twistor
  Space Actions}},  \href{http://arxiv.org/abs/1604.08209}{{\tt 1604.08209}}.

\bibitem{Eastwood:2002su}
M.~G. Eastwood, \emph{{Higher symmetries of the Laplacian}},
  \href{http://dx.doi.org/10.4007/annals.2005.161.1645}{\emph{Annals Math.}
  {\bf 161} (2005) 1645--1665},
  [\href{http://arxiv.org/abs/hep-th/0206233}{{\tt hep-th/0206233}}].

\bibitem{Vasiliev:2003ev}
M.~A. Vasiliev, \emph{{Nonlinear equations for symmetric massless higher spin
  fields in (A)dS(d)}},
  \href{http://dx.doi.org/10.1016/S0370-2693(03)00872-4}{\emph{Phys. Lett.}
  {\bf B567} (2003) 139--151}, [\href{http://arxiv.org/abs/hep-th/0304049}{{\tt
  hep-th/0304049}}].

\bibitem{Klebanov:2002ja}
I.~R. Klebanov and A.~M. Polyakov, \emph{{AdS dual of the critical O(N) vector
  model}}, \href{http://dx.doi.org/10.1016/S0370-2693(02)02980-5}{\emph{Phys.
  Lett.} {\bf B550} (2002) 213--219},
  [\href{http://arxiv.org/abs/hep-th/0210114}{{\tt hep-th/0210114}}].

\bibitem{Beccaria:2015vaa}
M.~Beccaria and A.~Tseytlin, \emph{{On higher spin partition functions}},
  \href{http://dx.doi.org/10.1088/1751-8113/48/27/275401}{\emph{J.Phys.} {\bf
  A48} (2015) 275401}, [\href{http://arxiv.org/abs/1503.08143}{{\tt
  1503.08143}}].

\bibitem{Giombi:2013yva}
S.~Giombi, I.~R. Klebanov, S.~S. Pufu, B.~R. Safdi and G.~Tarnopolsky,
  \emph{{AdS Description of Induced Higher-Spin Gauge Theory}},
  \href{http://dx.doi.org/10.1007/JHEP10(2013)016}{\emph{JHEP} {\bf 1310}
  (2013) 016}, [\href{http://arxiv.org/abs/1306.5242}{{\tt 1306.5242}}].

\bibitem{Tseytlin:2013jya}
A.~A. Tseytlin, \emph{{On partition function and Weyl anomaly of conformal
  higher spin fields}},
  \href{http://dx.doi.org/10.1016/j.nuclphysb.2013.10.009}{\emph{Nucl.Phys.}
  {\bf B877} (2013) 598--631}, [\href{http://arxiv.org/abs/1309.0785}{{\tt
  1309.0785}}].

\bibitem{Beccaria:2014xda}
M.~Beccaria and A.~A. Tseytlin, \emph{{Higher spins in AdS$_{5}$ at one loop:
  vacuum energy, boundary conformal anomalies and AdS/CFT}},
  \href{http://dx.doi.org/10.1007/JHEP11(2014)114}{\emph{JHEP} {\bf 1411}
  (2014) 114}, [\href{http://arxiv.org/abs/1410.3273}{{\tt 1410.3273}}].

\bibitem{Giombi:2014iua}
S.~Giombi, I.~R. Klebanov and B.~R. Safdi, \emph{{Higher Spin
  AdS$_{d+1}$/CFT$_d$ at One Loop}},
  \href{http://dx.doi.org/10.1103/PhysRevD.89.084004}{\emph{Phys.Rev.} {\bf
  D89} (2014) 084004}, [\href{http://arxiv.org/abs/1401.0825}{{\tt
  1401.0825}}].

\bibitem{Joung:2015eny}
E.~Joung, S.~Nakach and A.~A. Tseytlin, \emph{{Scalar scattering via conformal
  higher spin exchange}},
  \href{http://dx.doi.org/10.1007/JHEP02(2016)125}{\emph{JHEP} {\bf 02} (2016)
  125}, [\href{http://arxiv.org/abs/1512.08896}{{\tt 1512.08896}}].

\bibitem{Jacob:1959at}
M.~Jacob and G.~C. Wick, \emph{{On the general theory of collisions for
  particles with spin}},
  \href{http://dx.doi.org/10.1016/0003-4916(59)90051-X}{\emph{Annals Phys.}
  {\bf 7} (1959) 404--428}.

\bibitem{Freedman:1991tk}
D.~Z. Freedman, K.~Johnson and J.~I. Latorre, \emph{{Differential
  regularization and renormalization: A New method of calculation in quantum
  field theory}},
  \href{http://dx.doi.org/10.1016/0550-3213(92)90240-C}{\emph{Nucl. Phys.} {\bf
  B371} (1992) 353--414}.

\bibitem{Metsaev:2007fq}
R.~R. Metsaev, \emph{{Ordinary-derivative formulation of conformal low spin
  fields}}, \href{http://dx.doi.org/10.1007/JHEP01(2012)064}{\emph{JHEP} {\bf
  01} (2012) 064}, [\href{http://arxiv.org/abs/0707.4437}{{\tt 0707.4437}}].

\bibitem{Metsaev:2007rw}
R.~R. Metsaev, \emph{{Ordinary-derivative formulation of conformal totally
  symmetric arbitrary spin bosonic fields}},
  \href{http://dx.doi.org/10.1007/JHEP06(2012)062}{\emph{JHEP} {\bf 06} (2012)
  062}, [\href{http://arxiv.org/abs/0709.4392}{{\tt 0709.4392}}].

\bibitem{Elvang:2013cua}
H.~Elvang and Y.-t. Huang, \emph{{Scattering Amplitudes}},
  \href{http://arxiv.org/abs/1308.1697}{{\tt 1308.1697}}.

\bibitem{Gleisberg:2003ue}
T.~Gleisberg, F.~Krauss, K.~T. Matchev, A.~Schalicke, S.~Schumann and G.~Soff,
  \emph{{Helicity formalism for spin-2 particles}},
  \href{http://dx.doi.org/10.1088/1126-6708/2003/09/001}{\emph{JHEP} {\bf 09}
  (2003) 001}, [\href{http://arxiv.org/abs/hep-ph/0306182}{{\tt
  hep-ph/0306182}}].

\bibitem{Datta:2004mr}
A.~Datta, E.~Gabrielli and B.~Mele, \emph{{More on violation of
  angular-momentum selection rules in quantum gravity}},
  \href{http://arxiv.org/abs/hep-ph/0406149}{{\tt hep-ph/0406149}}.

\bibitem{koekoek1996askey}
R.~Koekoek and R.~F. Swarttouw, \emph{{The Askey-scheme of hypergeometric
  orthogonal polynomials and its q-analogue}}, {\emph{arXiv preprint
  math/9602214} (1996) }.

\bibitem{eden1967high}
R.~J. Eden, \emph{{High energy collisions of elementary particles}}.
\newblock Cambridge UP, 1967.

\bibitem{collins}
P.~Collins and E.~J. Squires, \emph{{Regge poles in particle physics}}.
\newblock Springer, 1968.

\bibitem{Ponomarev:2016jqk}
D.~Ponomarev and A.~A. Tseytlin, \emph{{On quantum corrections in higher-spin
  theory in flat space}},
  \href{http://dx.doi.org/10.1007/JHEP05(2016)184}{\emph{JHEP} {\bf 05} (2016)
  184}, [\href{http://arxiv.org/abs/1603.06273}{{\tt 1603.06273}}].

\bibitem{Fradkin:1981iu}
E.~S. Fradkin and A.~A. Tseytlin, \emph{{Renormalizable asymptotically free
  quantum theory of gravity}},
  \href{http://dx.doi.org/10.1016/0550-3213(82)90444-8}{\emph{Nucl. Phys.} {\bf
  B201} (1982) 469--491}.

\bibitem{Fradkin:1981jc}
E.~S. Fradkin and A.~A. Tseytlin, \emph{{One Loop Beta Function in Conformal
  Supergravities}},
  \href{http://dx.doi.org/10.1016/0550-3213(82)90481-3}{\emph{Nucl. Phys.} {\bf
  B203} (1982) 157--178}.

\bibitem{Lee:1982cp}
S.~C. Lee and P.~van Nieuwenhuizen, \emph{{Counting of States in Higher
  Derivative Field Theories}},
  \href{http://dx.doi.org/10.1103/PhysRevD.26.934}{\emph{Phys. Rev.} {\bf D26}
  (1982) 934}.

\bibitem{Riegert:1984hf}
R.~J. Riegert, \emph{{The particle content of linearized conformal gravity}},
  \href{http://dx.doi.org/10.1016/0375-9601(84)90648-0}{\emph{Phys. Lett.} {\bf
  A105} (1984) 110--112}.

\bibitem{gustavsson2001some}
J.~Gustavsson, \emph{{Some sums of Legendre and Jacobi polynomials}},
  {\emph{Mathematica Bohemica} {\bf 126} (2001) 141--149}.

\bibitem{Dona:2015tra}
P.~Dona, S.~Giaccari, L.~Modesto, L.~Rachwal and Y.~Zhu, \emph{{Scattering
  amplitudes in super-renormalizable gravity}},
  \href{http://dx.doi.org/10.1007/JHEP08(2015)038}{\emph{JHEP} {\bf 08} (2015)
  038}, [\href{http://arxiv.org/abs/1506.04589}{{\tt 1506.04589}}].

\bibitem{Stelle:1977ry}
K.~S. Stelle, \emph{{Classical Gravity with Higher Derivatives}},
  \href{http://dx.doi.org/10.1007/BF00760427}{\emph{Gen. Rel. Grav.} {\bf 9}
  (1978) 353--371}.

\bibitem{Dolan:2008gc}
L.~Dolan and J.~N. Ihry, \emph{{Conformal Supergravity Tree Amplitudes from
  Open Twistor String Theory}},
  \href{http://dx.doi.org/10.1016/j.nuclphysb.2009.04.003}{\emph{Nucl. Phys.}
  {\bf B819} (2009) 375--399}, [\href{http://arxiv.org/abs/0811.1341}{{\tt
  0811.1341}}].

\bibitem{Berkovits:2004jj}
N.~Berkovits and E.~Witten, \emph{{Conformal supergravity in twistor-string
  theory}}, \href{http://dx.doi.org/10.1088/1126-6708/2004/08/009}{\emph{JHEP}
  {\bf 08} (2004) 009}, [\href{http://arxiv.org/abs/hep-th/0406051}{{\tt
  hep-th/0406051}}].

\bibitem{Adamo:2012xe}
T.~Adamo and L.~Mason, \emph{{Twistor-strings and gravity tree amplitudes}},
  \href{http://dx.doi.org/10.1088/0264-9381/30/7/075020}{\emph{Class. Quant.
  Grav.} {\bf 30} (2013) 075020}, [\href{http://arxiv.org/abs/1207.3602}{{\tt
  1207.3602}}].

\bibitem{Adamo:2013tja}
T.~Adamo and L.~Mason, \emph{{Conformal and Einstein gravity from twistor
  actions}},
  \href{http://dx.doi.org/10.1088/0264-9381/31/4/045014}{\emph{Class. Quant.
  Grav.} {\bf 31} (2014) 045014}, [\href{http://arxiv.org/abs/1307.5043}{{\tt
  1307.5043}}].

\bibitem{Maldacena:2011mk}
J.~Maldacena, \emph{{Einstein Gravity from Conformal Gravity}},
  \href{http://arxiv.org/abs/1105.5632}{{\tt 1105.5632}}.

\bibitem{Maldacena:2011jn}
J.~Maldacena and A.~Zhiboedov, \emph{{Constraining Conformal Field Theories
  with A Higher Spin Symmetry}},
  \href{http://dx.doi.org/10.1088/1751-8113/46/21/214011}{\emph{J. Phys.} {\bf
  A46} (2013) 214011}, [\href{http://arxiv.org/abs/1112.1016}{{\tt
  1112.1016}}].

\bibitem{Boulanger:2013zza}
N.~Boulanger, D.~Ponomarev, E.~D. Skvortsov and M.~Taronna, \emph{{On the
  uniqueness of higher-spin symmetries in AdS and CFT}},
  \href{http://dx.doi.org/10.1142/S0217751X13501625}{\emph{Int. J. Mod. Phys.}
  {\bf A28} (2013) 1350162}, [\href{http://arxiv.org/abs/1305.5180}{{\tt
  1305.5180}}].

\bibitem{Stanev:2013qra}
Y.~S. Stanev, \emph{{Constraining conformal field theory with higher spin
  symmetry in four dimensions}},
  \href{http://dx.doi.org/10.1016/j.nuclphysb.2013.09.002}{\emph{Nucl. Phys.}
  {\bf B876} (2013) 651--666}, [\href{http://arxiv.org/abs/1307.5209}{{\tt
  1307.5209}}].

\bibitem{Alba:2015upa}
V.~Alba and K.~Diab, \emph{{Constraining conformal field theories with a higher
  spin symmetry in d> 3 dimensions}},
  \href{http://dx.doi.org/10.1007/JHEP03(2016)044}{\emph{JHEP} {\bf 03} (2016)
  044}, [\href{http://arxiv.org/abs/1510.02535}{{\tt 1510.02535}}].

\end{thebibliography}\endgroup
\bibliographystyle{JHEP}

\end{document}